\documentclass[11pt,a4paper]{article}

\usepackage{polski}
\usepackage[english]{babel}

\usepackage{jheppub}
\usepackage{amsthm,amssymb,amsmath,epic,float}
\usepackage{rotating,indentfirst,array,varioref}
\usepackage{appendix,tikz,mathtools}
\usepackage{setspace}
\usepackage{longtable}
\usepackage{tikz}
\usetikzlibrary{decorations.pathreplacing}
\usetikzlibrary{calc}
\usepackage{enumitem}
\usepackage{hyphenat}
\usetikzlibrary{positioning}
\usepackage{graphicx}  
\usepackage{adjustbox}

\usepackage{arydshln}
\usepackage{cancel}

\usepackage{subfig}
\usepackage{comment}
\usetikzlibrary{intersections}
\pgfdeclarelayer{background}
\pgfsetlayers{background,main}

\usepackage{tikz,pgf}
\usetikzlibrary{shapes}
\usetikzlibrary{calc}
\usetikzlibrary{decorations.pathmorphing}
\usetikzlibrary{decorations.pathreplacing,shapes.misc}
\usetikzlibrary{positioning}
\usetikzlibrary{arrows}
\usetikzlibrary{decorations.markings}
\usetikzlibrary{shadings}
\usetikzlibrary{intersections}

\def\be{\begin{equation}}
\def\ee{\end{equation}}

\def\be{\begin{equation}}
\def\en{\end{equation}}
\def\ber{\begin{eqnarray}}
\def\enr{\end{eqnarray}}

\newcommand{\pd}{\partial}

\newcommand{\br}[1]{{\overline{#1}}}

\def\<{\left(}
\def\>{\right)}

\usepackage{jheppub}
\makeatletter
\def\@fpheader{\vspace{-.1cm}}
\makeatother

\title{\boldmath $x-y$ swap for $(2,2p+1)$ minimal string}

\author[a,b]{Aleksandr~Artemev}

\affiliation[a]{Landau Institute for Theoretical Physics, 142432, Chernogolovka, Russia}
\affiliation[b]{Skolkovo Institute of Science and Technology, 121205, Moscow, Russia}

\emailAdd{artemev.aa@phystech.edu}

\abstract{We continue the study of 2D gravity --- ``matrix model'' duality on the example of $(2,2p+1)$ minimal string. We propose a reformulation of the duality, related to a more conventional one by ``$x-y$ swap'' in the language of topological recursion. This formulation elucidates some conceptual and technical difficulties in the dictionary of the duality and relation to other examples. In particular, it allows to circumvent the necessity to use ``resonance transformations'', that were previously introduced to match the worldsheet and ``matrix model'' correlators, and the expressions for minimal string amplitudes in this approach are reminiscent of the ones obtained recently for ``complex Liouville string'' theory. Using this new approach, we also formulate a conjecture on how one can compute amplitudes with operators other than tachyons in the dual theory.}


\begin{document} 
\maketitle
\flushbottom
\section{Introduction}
Minimal string theory (MS) is defined by its worldsheet CFT that consists of $(p,q)$ Virasoro minimal model, BRST ghosts and Liouville CFT. It is one of the oldest studied examples of non-critical string theory. Due to exact solvability of the constituent CFTs and simple description of BRST cohomology in this theory perturbative string amplitudes, at least for low genus and number of insertions, can be computed exactly. One of its intriguing features is a conjectural dual description in terms of matrix models. For $(2,2p+1)$ series, which we focus on in this work, the dictionary of the duality on the level of correlators was developed over the years, culminating in \cite{belzam2009}, while for the general $(p,q)$ matter theory it is still not well understood at the moment. Recently similar matrix model descriptions were developed for different related models of 2D gravity, where matter theories are also governed by Virasoro symmetry: these are Virasoro minimal string (VMS, \cite{Collier:2023cyw}) and complex Liouville string (CLS, \cite{ Collier:2024worldsheet, Collier:2024matrixmodel}).

An important property characterizing the dual description is the recurrent relations for correlators, known as topological recursion (TR), which in matrix model roughly corresponds to $1/N$ expansion of loop equations. Topological recursion has been thoroughly studied on its own in the mathematical literature and is known to have a lot of interesting properties. For example, TR data can be expressed in terms of tautological intersection theory on the moduli space of stable curves \cite{Eynard:2011sc}, a connection that bridges the gap between TR and worldsheet interpretation in terms of ``Euler characteristic of bundle of conformal blocks'' for VMS.

Another one of these properties, studied in a different language in older works \cite{Kharchev:1993as, KawaiFuk} and recently investigated in more generality \cite{Alexandrov:2022ydc}, is the transformation of TR data under the so-called ``$x-y$ swap''. A relevant example of a system of correlators satisfying TR to have in mind are resolvent correlators in (e.g. hermitian) two-matrix models, where one integrates over matrices $X$ ad $Y$ with the weight of the form $\exp (-V(X,Y))$, $V(X,Y) =\text{tr\,}\left(V_1(X) + V_2(Y) + XY \right)$. In this case, ``$x-y$ swap'' transformation relates the correlators of resolvents that depend on $X$ and $Y$ in this model. Putative dual of minimal string can be described in this way (more precisely, as a ``double-scaling limit'' of such models). 

For $(2,2p+1)$ series, it reduces to a one-matrix model, if we are only interested in the observables that depend on $X$ (one can think that in thhe two-matrix model $V_2(Y)$ is quadratic and $Y$ can be integrated out). This is the language in which the duality is usually formulated. However, we would like to argue that for comparison with the worldsheet it might be more natural to think of ``$Y$-dependent'' observables instead. The main proposition of the current paper is that the map between MS worldsheet correlators and matrix model one, previously described in somewhat obscure terms, can be much more easily formulated using TR after ``$x-y$ swap''.

This paper is organized as follows. In section \ref{sec-worldsheet} we introduce notations and definitions and review known facts from worldsheet formulation of the theory, including some results following from ``higher equations of motion''. Section \ref{sec-matrix} is devoted to description of the relevant TR data and the dictionary proposed in \cite{belzam2009} to connect it to the worldsheet. In section \ref{sec-xyswap} we formulate our main proposal, providing and comparing explicit formulas for MS correlators in the new approach and briefly discuss its relations to other known 2D gravity theories. Finally, in section \ref{sec-gr-corr} we formulate a preliminary conjecture on how ``ground ring'' physical operators appear on the ``matrix model'' side in this language. Previously, the dictionary of the correspondence was only formulated for a class of ``tachyon'' operators, so we believe that the possibility of this extension is interesting. 

\section{Results and facts about worldsheet theory} \label{sec-worldsheet}
\subsection{Definitions and notations}
The worldsheet CFT of minimal string is a direct product of CFT minimal model, Liouville theory and BRST-ghost $BC$-system.

Central charge of Liouvile theory is parametrized as $c \equiv 1 +6 Q^2 \equiv 1 + 6 (b+b^{-1})^2$. The fact that total central charge is zero relates $b$ and minimal model parameters as $b = \sqrt{\frac{2}{2p+1}}$.

Primary fields of Liouville theory exist for any conformal dimension $\Delta = \br{
\Delta
}$. We will denote them in several different ways: $V_a$ is an operator of  dimension $\Delta = a(Q-a)$, with $a$ identified with $Q-a$; alternatively, one can parametrize it with Liouville momentum $P$: $\Delta = \frac{Q^2}{4}+P^2$, then $P$ and $-P$ are identified. The spectrum of Liouville theory, i.e. fields that appear in the OPE, have $P$ real. In minimal string, Liouville fields with $a = \frac{ (1-m) b^{-1} + (1-n) b}{2} \equiv a_{m,n} $, or $P  = i\frac{ m b^{-1} + n b}{2} \equiv P_{m,n} $ frequently appear. These fields will be denoted also by $V_{m,n}$. For positive integer $m,n$ these fields are degenerate on level $mn$: a null descendant of the form $D_{m,n}^{(L)} V_{m,n} = 0$ exists in the corresponding Verma module.

There is another parameter in Liouville theory, commonly called ``cosmological constant'' and denoted by $\mu$. Dependence on this parameter is essentially powerlike: $n$-point correlation function of primary fields $V_a$ in genus $g$
\begin{equation}
\langle V_{a_1}\dots V_{a_n} \rangle_g \sim \mu^{\frac{Q(1-g)-\sum a_i}{b}}.
\end{equation}
For calculations, one can put this parameter to any convenient value. We will refer to dependence on this parameter as ``KPZ scaling'' \cite{Knizhnik:1988ak} and $\frac{a_i}{b}$ as ``KPZ exponent'' of the operator $V_a$.

Central charge of the  $(2,2p+1)$  minimal model CFT is $c_M = 1- 6 (b^{-1}-b)^2$, with $b$ as above. The spectrum consists of a finite number of primary fields that will be denoted as $\Phi_{m,n}$. In $(2,2p+1)$ series $m=1$ and $n\in 1, \dots, p$ (sometimes it is convenient to extend this range to $n \in 1,\dots,2p$ with additional identification $n \equiv 2p+1-n$). These fields are degenerate on level $n$: $D_{1,n}^{(M)} \Phi_{1,n} = 0$.

Explicit form of structure constants in Liouville and minimal model CFTs will not be used, so we will not provide them here. See \cite{zamzam1996, Zamolodchikov:2005fy} for explicit expressions.

Finally, the ghosts are a free fermionic (with anticommuting fields) CFT that has two holomorphic fundamental fields $B,C$ with conformal dimensions $(2,0)$ and $(-1,0)$ respectively and two antiholomorphic $\br{B}, \br{C}$. The central charge of ghost CFT is $-26$. In the context of string theory, on nontrivial closed Riemann surface (e.g. torus) they are to be quantized with periodic boundary conditions along the cycles (see e.g. \cite{polchinski_1998}).

As usual in BRST quantization of string theory, there is a nilpotent ($Q_{\text{BRST}}^2 = 0$) BRST-symmetry with generator
\begin{equation}
Q_{\text{BRST}} \equiv \mathcal{Q} + \br{\mathcal{Q}};\,\mathcal{Q} \equiv \oint dz\, \left(C (T^L + T^M) + :C \pd C B: \right)
\end{equation}
and an analogous expression for antiholomorphic charge $\br{\mathcal{Q}}$. Cohomology of $Q_{\text{BRST}}$ defines physical operators, from which the string amplitudes are built. 

String amplitudes are integrals over the moduli space of conformal structures (or complex structures) $\mathcal{M}_{g,n}$ of certain worldsheet CFT correlators. For them to be non-zero, there should be an appropriate number of ghost insertions depending on the genus of the worldsheet; in particular, ghost number anomaly equation $N_C - N_B = 3-3g$ should be satisfied. While correlator in each constituent CFT depends explicitly on the metric on a surface (not only on its conformal class), zero total central charge and BRST-closedness of physical operators allow to construct a well-defined form on the moduli space from their proper combination. 

\subsection{Operators: tachyons and ground ring}
The set of bulk physical operators (BRST-cohomology classes) in minimal string theory was described by Lian and Zuckerman \cite{lianzuck}. There is a discrete family of them, one for each Liouville momentum $P_{1,-n};\,n\in \mathbb{Z},\, n \text{ mod }(2p+1) \neq 0$ (each operator is ``built'' from the corresponding Liouville field/its descendants and something from the minimal model). The ghost number of this operator is determined by $n$.

We will be interested in two classes of operators. The simplest ones are of ghost number $1$ and commonly referred to as ``tachyons''. They are obtained first by ``dressing'' minimal model primaries $\Phi_{1,n}$ with Liouville operators $V_a$ so that their total conformal dimension is (1,1). Such a dressing requires $a = a_{1,-n} = b \frac{n+1}{2}$. The obtained operators $U_{1,n} \equiv V_{1,-n}\Phi_{1,n}$ can then be dressed with $C, \br{C}$ ghosts to get a local BRST-invariant operator $W_{1,n}$ of dimension zero or integrated over the surface. We will also sometimes write $U_a,\,W_a$ for tachyon operators where dressing Liouville field is $V_a$.

The studies of the duality with matrix model mostly focused on the amplitudes involving only tachyons. The simplest one is a three-point function, that is obtained by multiplying Liouville and minimal model structure constants. It turns out \cite{Zamolodchikov:2005fy} that three-point amplitudes are simply equal to minimal model fusion number in proper normalization.

Tachyon amplitudes should be symmetric in the parameters of the correlator. In particular, for genus zero, when integrating over moduli space is equivalent to integration over $n-3$ cross-ratios, it should not depend on which $n-3$ fields are integrated.

A different nontrivial class of operators, having ghost number zero, is the so-called ``ground ring''. They were first discussed in \cite{wit1992}. These operators have the general form
\begin{equation}
O_{1,n} = H_{1,n} \br{H}_{1,n} \Theta_{1,n};\,\Theta_{1,n} \equiv V_{1,n} \Phi_{1,n}
\end{equation}
where $H$ is a polynomial of degree $n-1$ of ghost and Virasoro modes $L_{-k},M_{-k}$ (in Liouville and minimal model sector respectively). They are built from Virasoro descendants; the reason why they are, nevertheless, BRST-invariant, is because when acting with $\mathcal{Q}$ the result is a linear combination of null descendants $D^{(L)}_{m,n} V_{m,n}$ and the corresponding one in minimal model $D^{(M)}_{m,n} \Phi_{m,n}$(which we put to zero). This is the defining relation that allows to find the explicit form of $H_{1,n}$ case by case \cite{imbimbo1992} (although no general formula is known). The main example is $O_{1,2}$ with
\begin{equation}
H_{1,2} = L_{-1} - M_{-1} + b^2 :BC: \label{h12}   
\end{equation}
Ground ring operators are interesting due to the following properties: 
\begin{itemize}
    \item independence of the correlator on their position in the sense that
\begin{equation}
\pd H_{1,n} \Theta_{1,n} = \mathcal{Q} R_{1,n} \Theta_{1,n} + \mathcal{D}_{1,n} \Theta_{1,n}\label{posind}
\end{equation}
Here $R_{1,n}$ is another polynomial of Virasoro modes and ghosts and $\mathcal{D}_{1,n} = -(D^{(M)}_{1,n} + (-1)^n D^{(L)}_{1,n})$ 
\item very simple fusion among themselves
\begin{equation}
O_{m,n} O_{m',n'} = \sum \limits_{r = |m-m'|+1:2}^{m+m'+1} \sum \limits_{s = |n-n'|+1:2}^{n+n'+1} G_{r,s}^{(m,n)|(m',n')} O_{r,s} \label{ooope}
\end{equation}
and with ghost number 1 operators $W_a$:
\begin{equation}
O_{m,n} W_a = \sum \limits_{r = -m+1:2}^{m-1} \sum \limits_{s = -n+1:2}^{n-1} A_{r,s}^{(m,n)} (a) W_{a + \frac{rb^{-1} + s b}{2}} \label{owope}
\end{equation}
\end{itemize}
``$:2$'' denotes summation with step $2$ here and in what follows. The last two equations are valid in cohomology, i.e. up to $\mathcal{Q}$-exact terms; alternatively, their RHS gives a leading (independent of $x$) term in the OPE $O(x) O(0)$ and $O(x) W(0)$ respectively. The coefficients $G$ and $A$ in fact can be put to one if one considers renormalized operators $\mathcal{O}_{1,n} = \frac{O_{1,n}}{\Lambda_{1,n}},\,\mathcal{W}_{1,n} = \frac{W_{1,n}}{\mathcal{N}(a_{1,-n})}$ \cite{Belavin:2005jy}.

 Ground ring operators are closely associated with ``higher equations of motion'', an important structure in Liouville theory that allows for exact computation of some correlators in MS and related theories.  Ground ring appear naturally in such a derivation, which we will shortly review in the next section.
\subsection{Available results from higher equations of motion}
Higher equations of motion (HEM) were discovered in Liouville CFT in \cite{highereoms2004}. They concern the so-called ``logarithmic operators'', which are of the form
\begin{equation}
V'_a := \frac{1}{2} \frac{\pd}{\pd a} V_a
\end{equation}
We will denote by $V'_{m,n}$ such logarithmic operator evaluated at the point corresponding to degenerate dimension $a=a_{m,n}$. 
Logarithmic operators are not Virasoro primaries; however, higher equations of motion state that ``would-be null descendant'' of logarithmic operators behaves the same as $V_{m,-n}$ under the sign of correlation function (up to some multiple)
\begin{equation}
  D^{(L)}_{m,n}  \br{D}^{(L)}_{m,n}  V'_{m,n} = B_{m,n} V_{m,-n}
\end{equation}
This equality valid under the sign of correlation function up to contact terms, which are important if we want to integrate over some of operators' positions. HEM can also be formulated as kinematic properties, i.e. relations purely on conformal blocks. 

In gravity theories like the minimal string, they are applied as follows. Consider the operator $(\pd H_{1,n} - \mathcal{Q}R_{1,n})\Theta'_{1,n},\,\Theta'_{1,n} \equiv V'_{1,n} \Phi_{1,n}$. We get something similar to \eqref{posind}:
\begin{equation}
(\pd H_{1,n} - \mathcal{Q}R_{1,n})\Theta'_{1,n} = \mathcal{D}_{1,n} \Theta'_{1,n} + (\dots) \Theta_{1,n}
\end{equation}
Additional terms come from non-diagonal action of $L_0$ on logarithmic operators. But after acting with an antiholomorphic operator $(\br{\pd} \br{H}_{1,n} - \br{\mathcal{Q}R}_{1,n})$, it disappears, since $\mathcal{\br{D}}_{1,n} \Theta_{1,n} = 0$. In total we are left with
\begin{equation}
    (\pd H_{1,n} - \mathcal{Q}R_{1,n}) (\br{\pd} \br{H}_{1,n} - \br{\mathcal{Q}R}_{1,n})\Theta'_{1,n} = \mathcal{D}_{1,n} \br{\mathcal{D}}_{1,n} \Theta'_{1,n} = D^{(L)}_{1,n} \br{D}^{(L)}_{1,n} \Theta'_{1,n} \overset{\text{HEM}}{=} B_{1,n} U_{1,n}
\end{equation}
Thus, HEM allow to represent the integrated tachyon as a total derivative $\pd \br{\pd} O'_{1,n}$ up to BRST-exact terms. If all operators other than this tachyon are BRST-closed, at least formally BRST-exact contributions can be omitted; then, the integral reduces to a sum of boundary terms in the vicinity of other insertions. These boundary terms can be calculated, because only a finite number of contributions in the Liouville OPE are relevant for this. Namely, one can write

\begin{align}
&\mathcal{O}'_{1,n}(x) \mathcal{W}_a(0) = \log (x \br{x}) \sum \limits_{s=-n+1:2}^{n-1} q_{0,s}^{(1,n)}(a) \mathcal{W}_{a-s b/2} + \dots \label{gr-tach-ope}
 \\
&\mathcal{O}'_{1,k_2+1}(x)\mathcal{O}_{1,k_1+1}(0) = \log |x|^2 \sum \limits_{s=|k_2-k_1|:2}^{k_2+k_1} q_{0,s-k_1}^{(1,k_2+1)}(a_{1,k_1+1}) \mathcal{O}_{1,1+s}+ \dots \label{gr-gr-ope}
\end{align}
with
\begin{equation}
q_{r,s}^{(m,n)} \equiv |a +i P_{r,s} - \frac{Q}{2}| +i P_{m,n}
\end{equation}
For string amplitudes given by integral over low-dimensional moduli spaces HEM can be effectively applied. The first successful example of that was the tachyon four-point function in genus zero \cite{Belavin:2005jy, alesh2016}. We will not repeat the corresponding analysis; in proper normalization (up to the factor that only depends on central charge and leg-factors $\mathcal{N}$) it reads
\begin{equation} 
    \left\langle \int d^2x\,\mathcal{U}_{1,n}(x) \mathcal{W}_{1,k_1}(x_1)  \mathcal{W}_{1,k_2}(x_2) \mathcal{W}_{1,k_3}(x_3)\right\rangle \sim -2i n P_{1,n} +\sum \limits_{i=1}^3 \sum \limits_{s=-n+1:2}^{n-1} q_{0,s}^{(1,n)}(a_{1,-k_i}) \label{4-pf-HEM}
\end{equation}
This formula is applicable in a case when the parameters $k_i$ are ``generic''; this means that dimension of spaces of conformal blocks of the minimal model part of the correlator is equal to~$n$. Alternatively, as discussed in the original paper, it can be made sense of in the context of ``generalized minimal string''\footnote{Perhaps one should think of this as VMS with matter CFT ``completed with degenerate operators''. By this we mean the following: while the usual fields $\Phi_\alpha$ in timelike Liouville matter theory are not degenerate even at Kac values of dimensions $\Delta_{m,n}$, since structure constants obey the difference relations that would follow from BPZ null-descendant decoupling equation, one can consistently add ``genuine'' degenerate fields to the theory. Then, in the context of gravity one can compute amplitudes with dressed degenerate field and three usual ``generic'' ones. These correlators are different than the usual VMS ``quantum volumes''; presumably HEM should work for them.}, where we allow some of matter dimensions to have non-degenerate values. It is also applicable in the context of CLS \cite{Collier:2024worldsheet}. Using HEM in slightly different form, one can also compute torus one-point tachyon amplitudes in MS \cite{Artemev:2022sfi}. 

We will be also interested in the correlation function of four tachyons and one ground ring operator, for which HEM should be applicable. This will be discussed in section \ref{sec-gr-corr}.

\subsection{Discontinuity analysis}\label{disc-analysis}
In this section we would like to comment on an important property, coming from worldsheet formulation, that is expected from amplitudes in minimal string and related theories. This discussion is similar to the one in \cite{Collier:2024worldsheet}. 

Consider a sphere tachyon four-point amplitude for simplicity.  It is of the form
\begin{align*}
\int d^2 z\,&\int \limits_C\frac{dP}{4\pi}\,  C^L(P_1, P_2, P) C^L(-P,P_3,P_4) |\mathcal{F}^{L,b}(P,P_i|z)|^2 \times  \\
& \times \sum \limits_{\hat{P}} C^M(iP_1, iP_2, \hat{P}) C^M(iP_3, iP_4, \hat{P})  |\mathcal{F}^{M,ib}(\hat{P},iP_i|z)|^2 (z\br{z})^{P^2 + \hat{P}^2 - 1}
\end{align*}
where $\mathcal{F}$ is the four-point conformal block. While the focus is on the MS case, we will comment on 4-point correlator in VMS and CLS in parallel. The formula above can be applied to these cases as well with the suitable modification of what is meant by $\sum \limits_{\hat{P}}$.

It can be shown that for values of parameters appearing in MS the $z$ integral  always converges in the vicinity of the boundary of moduli space $(z \to 0,1, \infty)$. For particular $P_i$ integration contour $C$ is deformed from the real line and ``discrete terms'' appear, but the corresponding integrals are also convergent. There are possible issues with divergences when we continue away from this ``physical'' kinematics in the context of ``generalized minimal string''. 

In CLS and VMS, continuation away from ``physical values'' is associated with additional discrete terms in Liouville OPE with dimension less than $Q^2/4$, coming from the poles of structure constants $C^L$ crossing the integration contour. After integrating over $\hat{P}$, integral over $d^2z$ might cease to converge. 

To analytically continue to this regime, we can first perform an integral over $z$. Starting with a discrete term at intermediate momentum $P^*$ s.t. $\text{Re\,}P^{*2}>0$, if conformal blocks are expanded to leading order, integral over $z$ over the vicinity of $z=0$ gives
\begin{equation}
\int \limits_{|z|<\epsilon} d^2z\,|z|^{2(P^{*2} + \hat{P}^2 -1)} = \pi \frac{\epsilon^{2(P^{*2} + \hat{P}^2)}}{P^{*2} + \hat{P}^2}
\end{equation}
In other OPE channels, there are analogous singular contributions from other points at the boundary of moduli space $z=1, \infty$. Integral over the remaining part of the sphere is nonsingular as a function of $\hat{P}$. Analytic continuation of the resulting integrand (as a function of $\hat{P}$) then has a pole at $P^* = i\hat{P}$, with the calculable residue independent of the size $\epsilon$ of the disks we cut. We refer to these as ``on-shell poles'', since the intermediate states in Liouville + matter have total dimension $1$. The subleading terms in the OPE could give analogous poles only if we combine equal powers of $z$ and $\br{z}$ (for integral over polar angle to be non-zero). It turns out then that the corresponding coefficient in the product of conformal blocks cancels these poles. While this is not particularly relevant to this work, we provide a sketch of the simple proof of this statement based on Zamolodchikov recursion in Appendix \ref{app-on-shell}.

Now consider analytically continuing external parameters such that $P^*$ moves to the wedge of divergence. The on-shell pole will then drift in the $\hat{P}$ plane. We can approach some points on the integration contour in two directions, on one of which the residue at this pole will be picked on the way. We thus encounter a discontinuity in our amplitude equal to this residue. 

The situation now becomes different for the CLS and VMS cases. In the first case, one can show that the discontinuity is nonzero and proportional to $P^*$ times the residue of the product of 3-point amplitudes $A(P_1, P_2,P^*) A(P_3, P_4, P^*)$ \cite{Collier:2024worldsheet}.  For VMS, we expect amplitudes to be analytic and have no discontinuities. This appears as follows: if the spacelike Liouville structure constants have a pole at $P = P^*$ (residue in which gives a discrete term), timelike Liouville structure constants are zero at $\hat{P} = iP^*$. On-shell residue is proportional to timelike Liouville structure constants and thus is zero. 

For gMS we should switch the roles of Liouville and matter in the discussion above: we do not need to take a double residue, since intermediate dimensions in matter sector already take discrete values if at least one operator is degenerate. We can again analytically continue to some values of external parameters corresponding to divergent integral to see a discontinuity similar to CLS. The three-point amplitude is now equal to 1 (in ``generalized'' case, while in the ``physical'' one it is the fusion number).

Computing the residue (normalized by ``leg factors'' and Liouville partition function) in this case, as in CLS, gives
\begin{equation}
\frac{1}{Z_L\prod \limits_{j=1}^4 \mathcal{N}(Q/2 + iP_j)}\text{Res\,}_{P = i\hat{P}} \left(\dots \right) \sim  \hat{P} 
\end{equation}
Discontinuities linear in $P$ signal about non-analytic terms in the amplitude of the form $\sqrt{P^2}$. It was explored a little bit for CLS; for generalized MS we know that they also appear from HEM answer (previously we wrote them as $|P|$ to conform with earlier references). The outtake should be that the analysis of on-shell poles allows to predict some parts of the amplitude. We expect that this in some sense should be valid for genuine MS as well.

One might conjecture that this analysis should carry over to the case of higher-point amplitudes. Namely, there are discontinuities at certain external momenta for each one-node degeneration limit of the worldsheet $C$ ($C(g,n_1+n_2) \to C(g_1, n_1+1) \times C(g_2, n_2+1) $ or a non-separating one); the discontinuity is proportional to $P_{int}$ --- the intermediate momentum flowing through the node --- times the product of string amplitudes.

\section{Topological recursion for MS} \label{sec-matrix}
As mentioned in the introduction, conventionally the dual to $(p,q)$ minimal string is described in terms of (double-scaled) two-matrix models. While having this more specific framework is certainly important for studying and matching non-perturbative effects on two sides (see e.g. \cite{Eniceicu:2022dru, Gregori:2021tvs}),  we will use the language which is more convenient for our purposes of calculating perturbative string amplitudes --- the one of ``topological recursion'' \cite{Eynard:2007kz}. 
\subsection{MS spectral curve}
Input data for topological recursion is the so-called ``spectral curve''. For minimal string, it is a (degenerate) curve of genus zero defined parametrically by the following embedding in $\mathbb{C}^2$
\begin{equation}
\begin{cases}
x = 2 u_0\, T_2(z) \\
y = 2 u_0^{\frac{2p+1}{2}}\, T_{2p+1}(z) \\
\end{cases} \label{sc}
\end{equation}
where $T_k$ is the Chebyshev polynomial of the first kind and $u_0$ is a scaling parameter, related to cosmological constant $\mu$ in the worldsheet formulation; it will play a special role in what follows. This curve first appeared in \cite{Seiberg:2003nm}.

We will now describe topological recursion for genus zero spectral curves in some more generality. The 1-form $\omega_{0,1} = y\,dx$ and the bidifferential
\begin{equation}
    B(z_1, z_2) = \frac{dz_1 dz_2}{(z_1 - z_2)^2} \label{bergmank}
\end{equation}
define the following kernel for each ramification point $\zeta_m: dx(\zeta_m) = 0$ (we assume all zeroes of $dx$ are simple) as follows
\begin{equation}
K^{(m)}(z_0, z) = \frac{\int \limits_{\br{z}^{(m)}}^z B(z_0,\xi) d\xi}{2 (y(z) - y(\br{z}^{(m)})) dx} 
\end{equation}
Here by $\br{z}^{(m)}$ we denote the action of Galois involution on point $z$, i.e. it is a second solution of the equation $x(z) = x(\br{z}^{(m)})$ near $\zeta_m$. For the curve \eqref{sc}, there is only one ramification point $z=0$ and the involution acts simply via $\br{z}^{(1)} = - z$.

The system of $n$-differentials $\omega_{g,n}(z_1 \dots z_n) \equiv W_{g, n}(z_1, ..., z_n)\, dz_1 \dots dz_n$, which is the output of topological recursion, is now defined via
\begin{align} \label{toprecu}
\omega_{g,n+1}(z_1, \dots, z_n,z_{n+1}) =\,\sum \limits_m &\underset{z=\zeta_m}{\text{Res}}\;  K^{(m)}(z_{n+1},z) \left(\omega_{g-1,n+2}(z_1, \dots z_n,z,\br{z}^{(m)}) + \right. \nonumber \\
&\left. +\sum \limits_{g_1=0}^g \sum \limits_{J_1 +J_2 = \{z_1 \dots z_n \}}^{'} \omega_{g_1, |J|+1} (J_1, z) \omega_{g-g_1, |J_2|+1} (J_2, \br{z}^{(m)}) \right) 
\end{align}
Prime over the sum denotes that we omit the terms that would include $\omega_{0,1}$.

It is known that curves defined by a pair of polynomials in $\mathbb{C}^2$ define certain tau-functions $\mathcal{F}_g$ of (reduced) KP hierarchy (e.g. Krichever tau-function \cite{Marshakov:2024lvk, Krichever:1992qe} for $g=0$); KP times parametrize the moduli of the curve. $n$-differentials of TR, computed for a given spectral curve $(x,y)$, can be related to derivatives of this tau-function  at a certain point $t_k^0$, which can be computed as follows
\begin{equation}
 t^0_{k} = - \frac{1}{2}  \underset{z= \infty}{\text{Res}}\;  \left( x^{-1/2+ k-p} y\,dx \right)
\end{equation}
We use an unconventional way to enumerate them; a usual convention is $t_k = \frac{2p-2k+1}{2} t^{usual}_{2p-2k+1}$. We will only consider $t_k$ for $k=1, ..., p$ for now. One can check that even times $t_{2k}$ in this set are zero. The derivatives themselves are given by \cite{Alexandrov:2023jcj}
\begin{equation} \label{kp-deriv}
\frac{\pd^n \mathcal{F}_g}{\pd t_{k_1} \dots \pd t_{k_n}} \mid_{t_k = t_k^0} =\underset{z_1 = \infty}{\text{Res\,}}\dots \underset{z_n = \infty}{\text{Res\,}} \left(\omega_{g,n}(z_1, \dots, z_n) \prod \limits_{i=1}^n \frac{(-2)\,x^{p-k_i+1/2}(z_i)}{2p-2k_i+1} \right)
\end{equation}
This facts relates TR formulation to ``integrability'' and ``matrix model'' approaches to minimal string.
\subsection{Definition of correlators via resonance transformations} \label{reson-tr}
These derivatives, natural from integrability point of view, are not precisely what we are interested in. To compare the results with worldsheet CFT approach, it was proposed \cite{belzam2009} that we need to take derivatives with respect to different couplings $\tau_k$, which are related to $t$ via the so-called ``resonance transformations''.

Let us parametrize the times via $u_0$, $\tau_{k}$ as follows: 
\begin{equation} 
 t_{k} = (2p+1) u_0^{k+1} \sum \limits_{n=1}^{\lfloor \frac{k+1}{2}\rfloor} \sum \limits_{\substack{m_1 \dots m_n \geq 1 \\ \sum (m_l+1) = k+1}} \frac{\tau_{m_1} \dots \tau_{m_n}}{n!} \frac{(2p-2k+2n-3)!!}{(2p-2k-1)!!} \label{taudef}
\end{equation}
$\tau_1 = -\frac{1}{2}; \tau_2, \dots, \tau_{p} = 0$ are background values corresponding to \eqref{sc}. Changing $u_0$ and keeping $\tau_k$ the same corresponds to rescaling of $x,y$ (keeping the Chebyshev form of the spectral curve). Then, consider a singular in $u_0^2$ part\footnote{By ``singular in $u_0^2$'' we mean proportional to negative or odd positive powers of $u_0$}  of the following expression
\begin{equation} \label{agndef}
A^g_n(k_1, \dots, k_n) := u_0^{-\sum \limits_{i=1}^n (k_i+1)}  \frac{\pd^n \mathcal{F}}{\pd \tau_{k_1} \dots \pd \tau_{k_n}} \mid_{\tau_1 = -\frac{1}{2}; \tau_2, \dots, \tau_{p} = 0}
\end{equation}
The proposal of \cite{belzam2009} is that this singular part coincides with the tachyon correlation numbers $\langle \mathcal{W}_{1,k_1} \dots \mathcal{W}_{1,k_n} \rangle$ with parameters defined in the first section. The polynomial change of variables \eqref{taudef} was found by demanding that $A^g_{n, \text{ singular}}$ obey the corresponding CFT fusion rules. Explicit formulas were obtained up to five-point numbers in genus zero \cite{tarn2011} and two-point numbers in genus 1 \cite{beltar2010}.

Since our formulation using TR is a bit different from the one used in previous works, an illustration of this proposal for the three-point amplitudes in the case of $(2,11)$ model is given in appendix \ref{app:211}.

\subsection{$p$-deformed volumes}
``Resonance transformations'' together with ``Douglas string equation'' approach of \cite{belzam2009, beltar2010, tarn2011} in general expresses $A^g_{n,\text{ singular}}$ as piecewise-polynomial functions of $p,k_i$. There is a certain domain in parameter space where the expressions significantly simplify; this is the domain $k_i + k_j>p-3$ for all $i,j$. With these conditions, when applying resonance transformations one can effectively linearize them in $\tau_m$ and throw away higher derivatives $\frac{\pd^2 t_k}{\pd \tau_{k_1} \pd \tau_{k_2}}, \dots$. 

It turns out that in this domain $A^g_n(\vec{k})$ reduce to polynomials in $p$ and $k_i$, symmetric in $k_i$; in special normalization they give a certain deformation of Weil-Petersson volumes of moduli spaces of hyperbolic surfaces with geodesic boundaries $V_{g,n}^{p}(\vec{l})$, analytically continued to imaginary lengths $l = 2\pi i b^2(p+\frac{1}{2}-k)$; parameter of deformation is $b^2$. The first few examples are
\begin{align}
V_{0,4}^{p}(\vec{l}) = 2 \pi^2 + \frac{3}{2}\pi^2 b^4 + \frac{1}{2} \sum \limits_{i=1}^4 l_i^2  \label{v04} \\
 V_{0,5}^{p}(\vec{l}) =  10 \pi^4 + 14 \pi^4 b^4 + \frac{13}{2} \pi^4 b^8 + \left(3 \pi^2 + \frac{5}{2} \pi^2 b^4\right) \sum \limits_{i=1}^5 l_i^2  +
    \frac{1}{2} \sum \limits_{i<j}l_i^2 l_j^2  + \frac{1}{8}\sum \limits_{i=1}^5 l_i^4 \label{v05} \\
V_{1,1}^{p}(\vec{l}) =  \frac{\pi^2 }{12}+ \frac{1}{48} l_1^2  \label{v11} \\
V_{1,2}^{p}(\vec{l}) =  \frac{\pi ^4}{4} +\frac{5}{24} \pi ^4 b^4 +  \frac{7}{192}\pi ^4 b^8 + \left(\frac{1}{12} \pi ^2  + \frac{1}{24} \pi ^2 b^4 \right) \sum \limits_{i=1}^2 l_i^2 +\frac{1}{96} \sum \limits_{i< j}l_i^2 l_j^2 + \frac{1}{192} \sum \limits_{i=1}^2 l_i^4 \label{v12}
\end{align}
More explicit expressions for low $g,n$ can be found in \cite{Artemev:2024rck}. These expressions are (slightly) different then the ``quantum volumes'' $V_{g,n}^b(\vec{l})$ defined in \cite{Collier:2023cyw} for Virasoro minimal string. These objects $V_{g,n}^{p}$ were first described in \cite{mertens2021} and named ``$p$-deformed volumes''. There, they are defined as a certain integral transform of topological recursion $n$-differentials. Their relation to tachyon correlation numbers was later noted and explained in \cite{Artemev:2024rck}. Also, it was found that the integral transformation of \cite{mertens2021} can be greatly simplified to give
\begin{equation}
V_{g,n}^{p}( 2\pi i b^2(p+\frac{1}{2}-k_i)) \sim \underset{z_1 = 0}{\text{Res\,}}\dots \underset{z_n = 0}{\text{Res\,}}\;\left( \omega_{g,n}(z_1, \dots, z_n) \prod \limits_{i=1}^n \frac{T_{2k+1}(z_i)}{2k+1}]\right)
\end{equation}
which can be also thought of as inverse Laplace transform with respect to $s = \arcsin(z)$. One can equivalently take the residue at infinity, since these are the only two singularities of the defined $n$-differential. 

We will see later that $p$-deformed volumes do not only coincide with $A^g_n$ is certain domain, but in some sense are ``building blocks'' for tachyon amplitudes in a general case.
\section{$x-y$ swapped spectral curve and correlation numbers} \label{sec-xyswap}
The approach described in the previous section has several drawbacks. First, the resonance transformations themselves are somewhat unnatural and confusing from the point of view of matrix models/integrability/topological recursion. Second, the analytic formulas obtained using them and ``Douglas string equation'' are not very transparent from the CFT point of view. Finally, if we expect that the dual approach should carry information not only about tachyons correlators, but the ones including operators of other ghost numbers (e.g. ground ring), it is not quite clear how to incorporate them. Indeed, resonance transformations fix $t$ as polynomials of $\tau$ of the form \eqref{taudef} essentially because of KPZ scaling. On the other hand, ``times'' $\tau$ corresponding to e.g. ground ring operators would have non-positive KPZ dimension.

We now proceed to one of the main statements of this work: an alternative prescription to calculate $A^g_n$, which we believe to be able to (at least partially) solve all the caveats listed above. The proposal is as follows: the $x-y$ dual spectral curve with
\begin{equation}
\begin{cases}
\check{x} =2 u_0^{\frac{2p+1}{2}} T_{2p+1}(z) \\
\check{y} = 2 u_0\, T_{2}(z) \\
\end{cases} \label{sc2}
\end{equation}
and the same bidifferential $B$ provides a system of $n$-differentials $\check{\omega}_{g,n}$ such that the following transform\footnote{Note that degree of the $T$ polynomial is proportional to Liouville momentum for the corresponding tachyon: $P_{1,-k} = \frac{i b}{4}(2p+1-2k)$}
\begin{equation} \label{w-check-to-A}
\check{A}^g_n(k_1, \dots, k_n) = \underset{z_1 = \infty}{\text{Res}} \dots \underset{z_n = \infty}{\text{Res}} \left( \check{\omega}_{g,n}(z_1, \dots, z_n) \prod \limits_{i=1}^n \frac{T_{2(p-k_i)+1}(z_i)}{2(p-k_i)+1}  \right)
\end{equation}
coincides with $A^g_{n,\text{singular}}$ defined in \eqref{agndef} (up to overall normalization).

As of now this statement is a conjecture --- the general proof is lacking. However, this nontrivial coincidence was verified for all examples explicitly studied in the literature before; we will provide the answers in what follows. We note that a similar in spirit statement has already appeared in the literature (although in a different description of minimal gravity dual) in \cite{Belavin:2015ffa}, although the relation is not so clear and in this case agreement with the ``resonance transformations'' approach was not complete. 
\subsection{KP perspective on swapped spectral curve}
Let us first clarify why \eqref{w-check-to-A} and the dual spectral curve \eqref{sc2} is more natural from the integrable systems perspective. First, if we now calculate KP times according to the formula
\begin{equation}
 \check{t}^0_{k} = - \frac{1}{2}  \underset{z= \infty}{\text{Res}}\;  \left( x^{-(2p-2k+1)/(2p+1)} y\,dx \right)
\end{equation}
we get a much simpler answer then before: only two of the times $t_1$ and $t_{-1}$ are nonzero. Moreover, the transformation \eqref{w-check-to-A}, that formally is the same as the one for calculating $p$-deformed volumes from the spectral curve \eqref{sc}, is now much more clear: the following property of Chebyshev polynomials
\begin{equation}
(T_{2p+1}(z))^{\frac{2k+1}{2p+1}} = 2^{\frac{2(p-k)}{2p+1}} T_{2k+1}(z) + O(1/z),\,z\to\infty;\,k\leq 2p
\end{equation}
means that at least for $k=1 \dots p$ corresponding to tachyon operators, \eqref{w-check-to-A} actually computes the derivatives with respect to KP times
\begin{equation}
\frac{\pd^n \mathcal{\check{F}}_g}{\pd \check{t}_{k_1} \dots \pd \check{t}_{k_n}} \mid_{\check{t}_k = \check{t}_k^0} = \underset{z_i = \infty}{\text{Res\,}} \left(\check{\omega}_{g,n}(z_1, \dots, z_n) \prod \limits_{i=1}^n \frac{\check{x}^{\frac{2p-2k_i+1}{2p+1}}(z_i)}{2p-2k_i+1} \right)
\end{equation}
\subsection{Explicit expressions for tachyon correlators and checks at low $g,n$}
The recursion kernel for (\ref{sc2}) in the vicinity of each branching point $\zeta_m = \cos \frac{\pi m}{2p+1},\,m = 1 \dots 2p$ 
\begin{align}
&K^{(m)}(z_0, z) \frac{dz}{dz_0} = \frac{\frac{1}{z_0-z} - \frac{1}{z_0 - \br{z}^{(m)}}}{8 u_0^{p+3/2} (z^2 - \br{z}^{(m)2}) (2p+1) U_{2p}(z)} = \\
& =\frac{1}{8(2p+1) u_0^{p+3/2}} \frac{1}{z + \br{z}^{(m)}} \frac{1}{(z_0 - z) (z_0 - \br{z}^{(m)}) U_{2p}(z)} 
\end{align}
($\br{z}^{(m)}(z)$ is the local Galois involution near $\zeta_m$; we omit the dependence on $m$ in what follows for brevity). Dependence on the factor $(4u_0^{p+3/2})$ will be trivial for all correlators; from now on we will omit it. 

As a simplest example, consider the 3-point correlator at genus zero. 
\begin{align}
&\check{W}_{0,3}(z_0,z_1,z_2) = \frac{1}{dz_0 dz_1 dz_2}\sum \limits_{m=1}^{2p} \underset{z=\zeta_m}{\text{Res}}\,K^{(m)}(z_0,z) \left(\check{\omega}_{0,2}(\br{z},z_1) \check{\omega}_{0,2}(z,z_2) + \check{\omega}_{0,2}(\br{z},z_2) \check{\omega}_{0,2}(z,z_1) \right) = \nonumber \\
&= \sum \limits_{m=1}^{2p} \underset{z=\zeta_m}{\text{Res}}\, \frac{d\br{z}/dz}{2(2p+1)} \frac{1}{z + \br{z}} \frac{1}{(z_0 - z) (z_0 - \br{z}) U_{2p}(z)}  \left(\frac{1}{(z-z_1)^2 (\br{z}-z_2)^2} + \frac{1}{(z-z_2)^2 (\br{z}-z_1)^2} \right) = \nonumber\\
& = \sum \limits_{m=1}^{2p} \frac{-2}{4(2p+1)} \frac{1}{(z_0-\zeta_m)^2 (z_1-\zeta_m)^2 (z_2-\zeta_m)^2} \frac{1}{\zeta_m U'_{2p}(\zeta_m)}  =\nonumber \\
& = \sum \limits_{m=1}^{2p}\underbrace{\frac{1}{2(2p+1)^2} (-1)^{m} \frac{\sin^2 \frac{\pi m}{2p+1}}{\cos \frac{\pi m}{2p+1}}}_{\equiv c_m}  \frac{1}{(z_0-\zeta_m)^2 (z_1-\zeta_m)^2 (z_2-\zeta_m)^2} 
\end{align}
Now perform a linear transformation \eqref{w-check-to-A} with respect to each variable $z_0, z_1, z_2$. The result is
\begin{equation}
\check{A}^0_{3} (\vec{k}) =\sum \limits_{m=1}^{2p} \frac{(-1)^{m}}{2(2p+1)^2} \frac{\sin^2 \frac{\pi m}{2p+1}}{\cos \frac{\pi m}{2p+1}}  \frac{\prod \limits_{i = 0}^2 \sin \frac{\pi m (2(p-k_i)+1)}{(2p+1)}}{\sin^3 \frac{\pi m}{2p+1}} = \frac{1}{2(2p+1)} \cdot b^2 \sum \limits_{m=1}^{2p}   \frac{\prod \limits_{i = 0}^2 \sin \frac{2\pi m k_i}{(2p+1)}}{\sin \pi m b^2}
\end{equation}
From worldsheet point of view (or ``resonance transformations'' appoach, by construction) we know that three-point tachyon amplitude at genus zero should be proportional to the three-point fusion number in the minimal model sector, i.e. equal to either zero or one, up to normalization. We can recognize it in the last factor; the fusion number (up to a sign) emerges in the form known as Verlinde formula \cite{Verlinde:1988sn} for minimal model. A similar representation exists for fusion numbers, or dimensions of space of minimal model conformal blocks, for more insertion points and in higher genus; e.g. in genus zero
\begin{equation}\label{verl}
\mathcal{N}^{(0)}_{i_1 \dots i_k} = (-1)^{\sum \limits_l (i_l-1) }\,  b^2 \sum \limits_{m=1}^{2p} \frac{ \prod \limits_{l=1}^k \sin \pi m i_l b^2}{(\sin \pi m b^2)^{k-2}}
\end{equation}
The next one in terms of difficulty is 4-point correlator on the sphere.
The corresponding 4-differential reads
\begin{align}
&\check{W}_{0,4}(z_0,z_1,z_2,z_3) = \nonumber \\
&=\sum \limits_{i=1}^3 \left( \sum \limits_{\underset{m,n=1..2p}{n \neq m}} \frac{c_n c_m}{z_{0m}^2 z_{im}^2 z_{\bullet n}^2 z_{\circ n}^2 (\zeta_m -\zeta_n)^2} + \sum \limits_{m=1}^{2p} \frac{c_m}{24(2p+1)^2 \zeta_m(\zeta_m^2-1)} \frac{P^m_{0,4}(z_0,z_i)}{z_{0m}^2 z_{im}^2 \prod \limits_{j=0}^3 z_{jm}^2} \right)
\end{align}
where
\begin{align}
P^m_{0,4}(z_0,z_1) = (-1)^m &\left[(-12 s_m^4) \left[z_{1m}^2 +3z_{0m}^2 \right] + 12 \zeta_m s_m^2 z_{0m} z_{1m}(3 z_{0m} + z_{1m}) + \right. \nonumber \\
&\left.+ z_{0m}^2 z_{1m}^2 (3\zeta_m^2 - 8p(1+p) s_m^2)  \right]
\end{align}
\begin{equation}
z_{im} \equiv z_i - \zeta_m,\, s_m^2 \equiv 1 -\zeta_m^2
\end{equation}
and $\circ$ and $\bullet$ in the first term denote two arguments other than $0$ and $i$ from the set $\lbrace 0,1,2,3 \rbrace$. Then one can perform a ``Chebyshev transform'' \eqref{w-check-to-A}. It can be brought to a simpler form if we represent $(m,n)$ dependent kernels in the expressions above, appropriately symmetrized with respect to $m \leftrightarrow 2p+1-m$, as sums of the form $\sum \limits_{k=1}^{2p} \sin (\pi k m b^2) \sin (\pi k n b^2) f(k)$. This rewriting allows to associate a Verlinde-like factor to each summation variable. For our kernels it turns out that $f(k)$ are simple enough --- polynomial in $|p+1/2-k|$. E.g.\footnote{Note that the identity below is only valid for $m,n \in 1\dots 2p$. It is not true if we consider both sides as functions of real variables $m,n$, which complicates proof of such identities. }
\begin{align}
&\frac{1}{b^2}\sum \limits_{k=1}^{2p} \sin (\pi m k b^2) \sin (\pi n k b^2) G(k) = \nonumber \\
& =(-1)^{n+m} \left(\frac{\delta_{m,n}-\delta_{m,2p+1-n}}{8} \left(\frac{s_m^2}{\zeta_m^2} - \frac{\zeta_m^2}{s_m^2} \right) + \frac{\delta_{m \neq n} \delta_{m+n \neq 2p+1}}{2} \left(\frac{s_m s_n}{(\zeta_m - \zeta_n)^2} + \frac{s_m s_n}{(\zeta_m + \zeta_n)^2} \right) \right)
\end{align}
where 
\begin{equation} \label{propagator}
 G(m) \equiv   \left(\frac{b^4}{12} + \frac{2}{3} + b^4 (p+\frac{1}{2}-m)^2 - 2b^2 |p+\frac{1}{2}-m| \right)
\end{equation}
The conjectural 4-point tachyon correlator then gets a nice representation 
\begin{equation} \label{04-answer}
\check{A}^0_{4} (\vec{k}) = \frac{b^2}{8} \left[\frac{V^p_{0,4}(l_i = 4\pi b P_{1,-k_i}))}{2\pi^2} \mathcal{N}^{(0)}_{k_1 \dots k_4} + \sum \limits_{i=2,3,4} \sum \limits_{m=1}^{2p}  \left(\frac{1}{2} \mathcal{N}^{(0)}_{k_1 k_i m} \mathcal{N}^{(0)}_{m \bullet \circ}\right) G(m) \right]
\end{equation}
where $V_{0,4}^{p}$ is given in \eqref{v04}. It has a similar structure as HEM answer \cite{Belavin:2005jy}, although the ingredients are different.

Let us perform some checks that the correlator in this form agrees with $A^{0}_{4}$ calculated in previous works. First, $\check{A}^0_{4}$ manifestly satisfies the fusion rules --- it is zero if $\mathcal{N}^{(0)}_{k_1 \dots k_4}$ is zero. 

\begin{figure}
    \centering
    \includegraphics[width=0.4\linewidth]{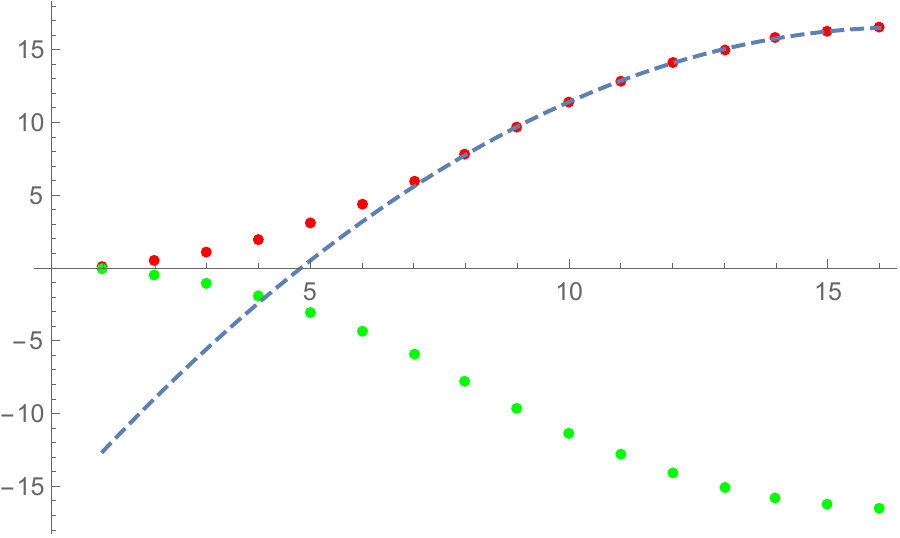}
  \includegraphics[width=0.4\linewidth]{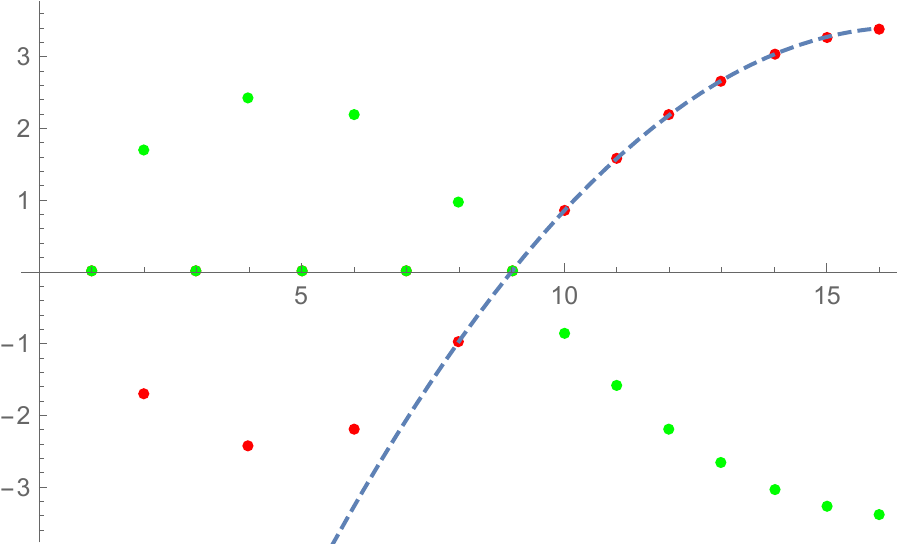}
  \caption{Two examples comparing the TR answer (red) and the one from ``resonance transformations'' approach in \cite{belzam2009} (green, plotted with the minus sign for more clear comparison) for 4-point amplitude with $p=16$: left --- $(k_1, k_1, p,p)$, right --- $(k_1, p/2,p/2,p/2)$ (parameter $k_1$ is on the $x$-axis). The dashed line is the corresponding $p$-deformed volume.}
    \label{fig:a04-check}
\end{figure}

 Second, consider the case when number of conformal blocks is equal to $k_1$. In this case, in every term in $\sum \limits_{i=2,3,4}$ fusion coefficients are non-zero only for $m = k_i-k_1+1 :2 :k_i+k_1-1$ (and ``reflected'', but to account for them we just erase the $1/2$ factor). The following sum can be taken:
\begin{equation}
\sum \limits_{m = k_i - k_1+1:2}^{k_i+k_1-1}( G(m) + 2b^2 |p+\frac{1}{2}-m| )= \frac{1}{12} k_1 \left(b^4 \left(4  k_1^2+12  k_i^2-3\right)-24 b^2  k_i +20\right)
\end{equation}
Then, summing  with $k_1 V_{0,4}$, the answer reduces to
\begin{equation}
\check{A}^0_4(\vec{k}) = \frac{b^4}{4} \left( k_1 (k_1 + p+ \frac{1}{2}) - \sum \limits_{i=2,3,4}\sum \limits_{m = k_i - k_1+1:2}^{k_i+k_1-1} |p+\frac{1}{2}-m| \right)
\end{equation}
This is the answer obtained for 4-point correlator using ``higher equations of motion'' (compare with \eqref{4-pf-HEM}, up to normalization, or with the expression in \cite{belzam2009}), which is valid precisely under the assumption $\mathcal{N}^{(0)}_{k_1 \dots k_4} = k_1$. Finally, we can perform some explicit numerical checks; figure \ref{fig:a04-check} illustrates that our formula coincides with the result in \cite{belzam2009}, in particular, reduces to $p$-deformed volume in the appropriate domain.

There is an alternative way to write this formula if instead of $p$-deformed volumes we express it using ``quantum volumes'' $V_{g,n}^b$ relevant to VMS and CLS\footnote{Here and in what follows we use a normalization in which $V_{g,n}^b$ reduce to WP volumes in the semiclassical limit $b \to 0$, different from the one in \cite{Collier:2023cyw} by a factor of $(8\pi^2 b^2)^{3g-3+n}$}:
\begin{equation} \label{04-answer-2}
\check{A}^0_{4} (\vec{k}) = \frac{b^2}{8} \left[\frac{V^b_{0,4}(l_i = 4\pi b P_{1,-k_i})}{2\pi^2} \mathcal{N}^{(0)}_{k_1 \dots k_4} + \sum \limits_{i=2,3,4} \sum \limits_{m=1}^{2p}  \left(\frac{1}{2} \mathcal{N}^{(0)}_{k_1 k_i m} \mathcal{N}^{(0)}_{m \bullet \circ}\right) \mathbf{G}(m) \right]
\end{equation}
The advantage is that there is a simpler expression for the modified ``propagator''
\begin{equation}
\mathbf{G}(m) = \left(\frac{2}{3} + b^4 (p+\frac{1}{2}-m)^2 - 2b^2 |p+\frac{1}{2}-m| \right) = 4 B_2\left(\frac{b^2}{2}|p+1/2-m| \right)
 \end{equation}
where $B_k(x)$ is the Bernoulli polynomial. In this form, MS answer practically coincides with CLS result (in section 5.4 of \cite{Collier:2024worldsheet}), up to a difference between the three-point amplitudes that enter the expression. Bernoulli polynomial in that context appeared from using an identity on polylogarithms.

Another example simple enough to be presented here is the 1-point correlation number in genus one. We have
\begin{equation}
\check{W}_{1,1}(z_0) = \sum \limits_{m=1}^{2p} \underset{z=\zeta_m}{\text{Res}} K^{(m)}(z_0,z) \check{W}_{0,2}(z,\br{z}) =  \sum \limits_{m=1}^{2p} (-1)^m \frac{P_{1,1}^{m}(z_{0m})}{48 (2p+1)^2 z_{0m}^4 s_m^2 \zeta_m}\,
\end{equation}
\begin{equation}
 P_{1,1}(x) = 3 s_m^4 +x^2 \left(2p(p+1) s_m^2 -3 \zeta_m^2\right) - 3x \zeta_m s_m^2
\end{equation}
and
\begin{equation}
\check{A}^1_1(k_0) = \sum \limits_{m=1}^{2p} \frac{\left(3 \zeta_m^2-2 k_0 (2p+1-k_0) s_m^2\right) \sin \left(\pi m b^2 k_0\right)}{48 (2 p+1)^2 \zeta_m s_m^3}
\end{equation}
Alternatively it can be written as
\begin{equation} \label{a11-diag}
\check{A}^1_1(k_0) = \frac{1}{4} \left(\mathcal{N}^{(1)}_{k_0} \frac{V_{1,1}^p (4\pi bP_{1,-k_0})}{2\pi^2} + \frac{1}{4} \sum \limits_{l=1}^{2p} \mathcal{N}^{(0)}_{k_0 l l} G(l) \right)
\end{equation}
with the fusion number at genus $1$ given by
\begin{equation}
\mathcal{N}^{(1)}_{k_0} = \frac{1}{2}(-1)^{k_0-1} \sum \limits_{m=1}^{2p} \frac{\sin \pi m k_0 b^2}{\sin \pi m b^2}
\end{equation}
Although it is not evident, for $k_0=1 \dots 2p$ \eqref{a11-diag} coincides with the simple answer obtained in \cite{Artemev:2022sfi, beltar2010}
\begin{equation}
A_1^1(k_0) = \frac{1}{24(2p+1)}k_0(2p+1-k_0)
\end{equation}

\subsection{``Feynman rules'' for $\check{A}^g_n$} \label{sec: feynman}
Based on the examples studied above, one can note that different terms in the expressions for $\check{A}_g^n$ can be associated to different stable graphs. These are the graphs $\Gamma$ enumerating all boundary strata in the moduli space of stable curves $\mathcal{\br{M}}_{g,n}$ (corresponding to different ways to degenerate the surface). They are defined as follows:
\begin{itemize}
    \item this is a connected graph with $n$ external ``legs'', vertices and the edges connecting some of the vertices. Legs correspond to marked points, vertices --- to smooth components of the normalization of degenerate surface and edges to the nodal points where degeneration happens.
    \item With each vertex $v$, a number $g_v \in \mathbb{Z}_{\geq 0}$ is associated which is a genus of the corresponding smooth component. For each $v$ a stability condition holds: $2g_v -2 + n_v  > 0$, where $n_v$ is the valence of $v$ including edges and legs.
    \item The total genus of the graph, defined as $\sum \limits_v g_v + h^1(\Gamma)$ ($h^1$ is the first Betti number), is equal to $g$.
\end{itemize}

This structure is typical of topological recursion. A similar one exists for CLS \cite{Collier:2024matrixmodel}.

It would be interesting to interpret stable graphs in MS case as ``Feynman diagrams'', associating specific analytic expressions to elements of these graphs so that we don't need to perform a (conceptually simple, but bulky) calculation for each new $\check{A}_{n}^g$. The simplest examples above suggest the following diagrammatic rules:
\begin{itemize}
    \item to each edge and leg associate a marking $k \in 1 ..p$ (for external legs these are the arguments of $\check{A}^g_n$)
    \item for each vertex $v$ with genus $g_v$ and $n_v$ incoming lines with markings $\vec{k}_v = k_1 \dots k_n$ multiply by $\frac{\mathcal{N}_{\vec{k}_v}^{(g_v)} V_{g,n}}{(2\pi^2)^{3g+n-3}}$
    \item for each internal edge (``propagator'') with marking $k_{int}$ multiply by $G(k_{int})$ and sum over $k_{int}$ from $1$ to $p$
    \item each graph is weighted by its symmetry factor $1/|\text{Aut}(\Gamma)|$ (e.g. the factor $1/2$ appears in the second term in \eqref{a11-diag})
\end{itemize}
However, the next examples that are presented in the appendix \ref{app:moreAgn} --- $(g,n) = (0,5)$ and $(g,n) = (1,2)$, although exhibit the expected structure as sum over ``stable graphs'', contain additional contributions that do not follow from the diagrammatic technique above. 

Based on these examples and a calculation in the limiting $p\to \infty$ case, to be discussed in the next section, we conjecture the following general expression:
\begin{align}
\check{A}^g_n(\vec{k}) =\sum \limits_\Gamma \frac{1}{|\text{Aut}(\Gamma)|} \sum \limits_{\vec{k_e};\, k_e=1}^{p}  \sum \limits_{\vec{n_e};\,n_e=1}^\infty \prod \limits_e \left[\frac{4^{n_e} B_{2n_e}(|bP_{1,-k_e}|)}{n_e!}\right] \prod \limits_{v} \mathcal{N}_{\vec{k}_v}^{(g_v)}  \times \nonumber \\
\times  \left(\prod \limits_e \left(\frac{-\pd}{4b^2\pd (P_{1,-k_e}^2)}\right)^{n_e-1} \prod \limits_{v}  \frac{ V_{g_v,n_v}^b(4\pi b P_{1,-\vec{k}_{v}})}{(2\pi^2)^{3g_v + n_v-3}}\right) \Bigg|_{P_e^2 = 0} 
\end{align}
The sum over $\Gamma$ includes all stable graphs of genus $g$ with $n$ legs with external markings $k_1 \dots k_n$. For each edge in the graph, a marking $k_e$ and a natural number $n_e$ are associated and the sum is over all these assignments. It is simpler to write this answer as a sum of tautological intersection numbers on $\br{\mathcal{M}}_\Gamma = \prod \limits_v \br{\mathcal{M}}_{g_v,n_v}$ --- moduli space of stable curves with the degenerations specified by stable graph $\Gamma$, as follows:
\begin{align}
\check{A}^g_n(\vec{k}) = \sum \limits_{\Gamma} \frac{1}{|\text{Aut}(\Gamma)|}\sum \limits_{\vec{k_e};\, k_e=1}^{p}  \prod \limits_{v} \mathcal{N}_{\vec{k}_v}^{(g_v)} \times \int \limits_{\br{\mathcal{M}}_\Gamma} \prod \limits_{v}e^{(1+b^4) \kappa_1  -  \sum \limits_{m \geq 1} \frac{B_{2m}}{(2m+1)!}(4b^2)^{2m} \kappa_{2m}}  \times \nonumber \\
\times\, e^{ \sum \limits_{i=1}^n (2bP_{1,-k_i})^2 \psi_i} \times \sum \limits_{\vec{n_e};\,n_e=1}^\infty\prod \limits_e  \frac{4^{n_e}B_{2n_e}(|bP_{1,-k_e}|) (-\psi_{e\bullet} - \psi_{e\circ})^{n_e-1}}{n_e!} \label{int-theor-conj}
\end{align}
In product over $v$ it is understood that each $\kappa$-class is an element in cohomology $H^{\bullet}(\br{\mathcal{M}}_{g_v,n_v})$; $\psi_{e\bullet}$ and $\psi_{e\circ}$ are $\psi$-classes, associated to two images of the node point, to which the edge $e$ corresponds, after normalization of the curve.  We will not give precise definitions of $\kappa$ and $\psi$-classes and refer to expositions in \cite{Zvonkine2012AnIT, Collier:2023cyw}.

 Higher terms in the series in $n_e$ (with $n_e >1$), of course, do not appear in the examples that we studied. However, in the JT limit (to be discussed in the next section) one can see how this series in Bernoulli polynomials appears explicitly.

Another way to rewrite the series in even Bernoulli polynomials in \eqref{int-theor-conj} is, for example, using an integral representation
\begin{align*}
   & \sum \limits_{n=1}^\infty   \frac{4^{n}B_{2n}(bP) u^{n-1}}{n!} = \int \limits_{-\epsilon -i \infty}^{-\epsilon+i \infty} \frac{dt}{2\pi i} \left(\frac{\pi}{\sin \pi t} \right)^2 \underbrace{  \frac{e^{4u (bP+t)^2}-1}{u}}_{f(P,t); f(-P,t) = f(P,-t)} = \\
    & = \frac{1}{2} \int \limits_{-\epsilon -i \infty}^{-\epsilon+i \infty} \frac{dt}{2\pi i} \left(\frac{\pi}{\sin \pi t} \right)^2 (f(P,t) + f(-P,t)) + \frac{1}{4} \underset{t=0}{\text{Res\,}} \left(\frac{\pi}{\sin \pi t} \right)^2  (f(P,t) - f(P,-t))
\end{align*}
It formally reduces to a sum of residues in the right half-plane (at each order in $u$ the series for Hurwitz zeta-function $\zeta(-2n+1,bP)$ is obtained, which reduces to Bernoulli polynomials). This series takes the form
\begin{align*}
&\sum \limits_{k = 1}^{\infty} \left(4( bP +k) e^{4u (bP+k)^2}  \right) + \sum \limits_{k = 1}^{\infty} \left(4(-bP +k) e^{4u (bP-k)^2}  \right) + 4 (bP) e^{4u (bP)^2} =\\
& = \sum \limits_{k \in \mathbb{Z}} \left(4 |bP +k| e^{4b^2 u (P+k/b)^2}  \right) 
\end{align*}
 This function is singular at $u \to 0$ and only has an asymptotic expansion in $u$, but the regular part of the corresponding asymptotic series coincides with the one that we started with. In this representation it is interesting that if we sum over $k\in \mathbb{Z}$ and over possible values of intermediate momentum $|bP_{1,-m}| = \frac{1}{2} - \frac{m}{2p+1},\,m=1 \dots p$ through the edge $e = (v_1,v_2)$, we get the following 
\begin{align}
&\sum \limits_{k=1}^p \sin (\pi k m_{v_1} b^2) \sin (\pi k m_{v_2} b^2) \sum \limits_{m \in \mathbb{Z}} \left(4 ||bP_{1,-k}| +m| e^{4b^2 u (|P_{1,-k}|+m/b)^2}  \right) = \nonumber\\
&=\sum \limits_{\underset{n \text{ mod }(2p+1) \neq 0}{n \in \mathbb{Z}}} 2b|P_{1,-n}| \sin (\pi n m_{v_1} b^2) \sin (\pi n m_{v_2} b^2) e^{4b^2 u |P_{1,-n}|^2} \label{expon-series}
\end{align}
The sine factors appear from Verlinde representation of fusion numbers appearing at the vertices $v_1, v_2$. First, note that this is just a ``discrete version'' of the ``propagator'' in CLS \cite{Collier:2024matrixmodel}, where the integral is replaced by a sum. Second, $P_{1,-n}$ in the total sum go through all the values in Lian-Zuckerman set, enumerating physical operators. The  restriction $n \text{ mod }(2p+1) \neq 0$ is in fact not required, since such terms automatically disappear because of sine factors from the fusion constants.  

If we substitute the series \eqref{expon-series} into \eqref{int-theor-conj}, we can perform integral over $\br{\mathcal{M}}_\Gamma$ explicitly in terms of ``quantum volumes''; recall that their definition in our normalization is
\begin{equation}
\frac{ V_{g,n}^b(4\pi b P_{1,-\vec{k}})}{(2\pi^2)^{3g + n-3}}  = \int \limits_{\br{M}_{g,n}} e^{(1+b^4) \kappa_1  -  \sum \limits_{m \geq 1} \frac{B_{2m}}{(2m+1)!}(4b^2)^{2m} \kappa_{2m} + \sum \limits_{i=1}^n (2bP_{1,-k_i})^2 \psi_i}  
\end{equation}
One then gets a very nice (albeit formal, since the series over $k_e$ is divergent and should be interpreted in zeta-regularization, as mentioned above) representation for $\check{A}^g_n$ 
\begin{align}
\check{A}^g_n(\vec{k}) = \sum \limits_{\Gamma} \frac{1}{|\text{Aut}(\Gamma)|}\sum \limits_{\vec{k_e}\in \mathbb{Z}}  \prod \limits_{v} \left( \mathcal{N}_{\vec{k}_v}^{(g_v)} \frac{ V_{g_v,n_v}^b(4\pi b P_{1,-\vec{k}_{v}})}{(2\pi^2)^{3g_v + n_v-3}} \right) \prod \limits_e 2b|P_{1,-k_e}| \label{int-theor-conj-final}
\end{align}
 
 The formula \eqref{int-theor-conj} should follow from the general expression for $\omega_{g,n}$ in terms of sum over stable graphs  \cite{Dunin-Barkowski:2012kbi} for some parametrization of the spectral curve, but such a derivation is technically involved. If this conjecture will be proven, we will have a  powerful method to obtain prediction for tachyon amplitudes in minimal string from dual approach.
\subsection{Comments on JT/Mirzakhani limit}
It is known that $p \to \infty$ limit of $(2,2p+1)$ minimal string is related to JT gravity. In particular, with appropriate scaling the spectral curve \eqref{sc} becomes the one for Mirzakhani topological recursion \cite{Mirzakhani:2006fta, saad2019jt}
\begin{equation} \label{sc-mirz}
\begin{cases}
   x(z) = z^2 \\
   y(z) = \frac{1}{4\pi} \sin 2\pi z \\
\end{cases}
\end{equation}
which calculates JT gravity partition functions with asymptotic boundaries of fixed renormalized lengths, or Weil-Petersson volumes. Minimal string tachyon correlators are known to reduce to ``WP volumes for hyperbolic surfaces with conical defects'' in this limit. To our knowledge, these objects were not studied comprehensively in the literature until recent works \cite{eberhardt20232d} (from a physical side) and \cite{anagnostou2023wp} (from more mathematical). Different checks of this proposal were performed before that in \cite{Artemev_2022, turiaci2021}. 

\begin{figure}
    \centering
    \includegraphics[width=0.4\linewidth]{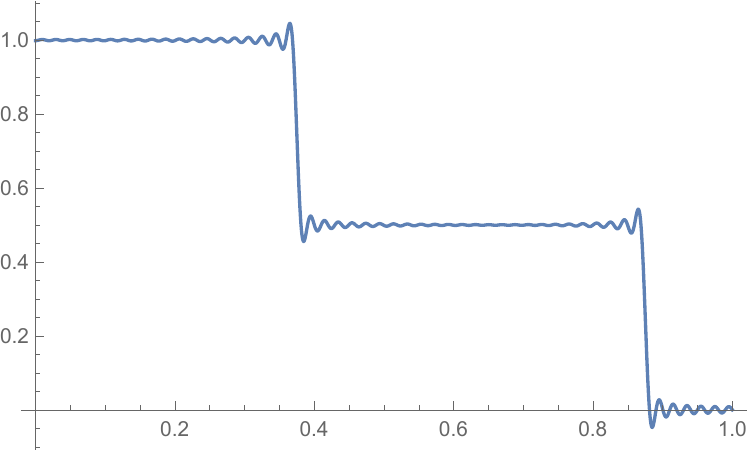}
  \caption{Example for 3-pt amplitude \eqref{a03-jt} with $\kappa_i =(\kappa_0, 3/8,1/4)$ (partially summed up to $m=100$)}
    \label{fig:JT-0-3}
\end{figure}

In \cite{eberhardt20232d}, in particular, an intersection-theoretic formula, inspired by the minimal string answers, as well as ``matrix integral'' formulation that can be used to compute these volumes were proposed. The results of the present paper suggests that there exists a different approach, in terms of $x -y$ swapped Mirzakhani spectral curve. We consider
\begin{equation} \label{sc-mirz-2}
\begin{cases}
   x^{JT}(z) = \frac{1}{4\pi} \sin 2\pi z \\
   y^{JT}(z) = z^2 \\
\end{cases}
\end{equation}
and the standard bidifferential; branching points for this curve are the zeroes of $\cos (2\pi z); z_m = \frac{1}{4} (2m+1), m \in \mathbb{Z}$. The local Galois involution $\br{z}(z)$ near $z_m$ is simply $2z_m - z$.
The kernel is then
\begin{equation}
K^{(m)}_{JT}(z_0, z) = \frac{\frac{1}{z_0-z} - \frac{1}{z_0 - \br{z}}}{(z^2 - \br{z}^2) \cos (2\pi z)} \frac{dz_0}{dz} = \frac{1}{2z_m (z_0 - z) (z_0 + z - 2z_m) \cos (2\pi z)} \frac{dz_0}{dz}
\end{equation}
Suitably scaled version of ``Chebyshev transform'' that should compute the volumes in this limit is the following:
\begin{equation} \label{w-to-A-JT}
A_{n}^{g\,JT}(\kappa_1,\dots, \kappa_n) = \underset{z_i= \infty}{\text{Res}}\left( \omega^{JT}_{g,n}(z_1, \dots, z_n) \prod \limits_{i=1}^n  \frac{\sin (2\pi \kappa_i z_i)}{2\pi\kappa_i}\right)
\end{equation}
$\kappa_i \in [0,1]$ corresponds to conical defect of angle deficit $2\pi (1-\kappa_i)$. Again let us start with the simplest example of three-point correlator in genus zero
\begin{align}
&W^{JT}_{0,3}(z_0,z_1,z_2)= \frac{1}{dz_0 dz_1 dz_2} \sum \limits_{m \in \mathbb{Z}} \underset{z=z_m}{\text{Res}} K^{(m)}_{JT}(z_0,z) \left(W_{0,2}(\br{z},z_1) W_{0,2}(z,z_2) + W_{0,2}(\br{z},z_2) W_{0,2}(z,z_1) \right) = \\
&= \sum \limits_{m \in \mathbb{Z}} \underset{z=z_m}{\text{Res}} \frac{(-1)}{2z_m} \frac{1}{(z_0 - z) (z_0 - 2z_m + z) \cos (2 \pi z)}  \left(\frac{1}{(z-z_1)^2 (\br{z}-z_2)^2} + \frac{1}{(z-z_2)^2 (\br{z}-z_1)^2} \right) = \\
& = -\sum \limits_{m \in \mathbb{Z}} \frac{(-1)^m}{2\pi z_m} \frac{1}{(z_0 - z_m)^2 (z_1 - z_m)^2 (z_2 - z_m)^2}
\end{align}
Transformation \eqref{w-to-A-JT} brings us to 
\begin{equation} \label{a03-jt}
   -A^{0\,JT}_{3}(\kappa_1, \kappa_2, \kappa_3) = \sum \limits_{m \in \mathbb{Z}} \frac{(-1)^m}{2\pi z_m}  \left(\prod \limits_{i=0}^2 \cos (2 \pi \kappa_i z_m) \right) 
\end{equation}
One can easily prove that this function is piecewise constant --- it is given as a sum of several terms which can be recognized as a Fourier series decomposition of a ``square wave'':
\begin{equation}
\sum \limits_{m \in \mathbb{Z}} \frac{2(-1)^m}{\pi(2m+1)} \cos \left(\frac{\pi \kappa}{2}(2m+1)\right) = 1-2\theta(1-|\kappa-2|),\,\kappa \in (0,4) 
\end{equation}
An example is given in fig. \ref{fig:JT-0-3}. For small enough $\kappa$ (``sharp defects'') $A^{0\,JT}_3$ is equal to 1, as expected of WP volume $V_{0,3}$. If $\sum (1-\kappa_i) < 2$, it is still nonzero for a while and takes value $1/2$. Geometrically this corresponds to the case when we are not anymore in the realm of hyperbolic geometry: because of Gauss-Bonnet theorem if the defects are ``blunt'' enough, the surface can only have a metric of constant positive curvature. From the point of view of MS $1/2$ appears because of averaging between answers from odd and even sectors (one of which is zero according to minimal model fusion rules). When ``Troyanov condition'' 
\begin{equation}
\sum \limits_{i=1}^3 (1-\kappa_i) > 2(1-\text{min}(\kappa_i))
\end{equation}
stops being satisfied, it is known that no metric of constant positive curvature exists at all \cite{mazzeo2015teichmuller}. This is precisely when three-point amplitude \eqref{a03-jt} becomes zero. It is interesting that these correlators compute some characteristics of both hyperbolic and ``euclidean de Sitter''-type geometries (probably this is related to ``anti-sphere'' complex saddle of semiclassical Liouville path integral that exists for these values of parameters \cite{Harlow_2011})

For the reference let us also provide an expression for four-point correlator in genus zero that follows from \eqref{sc-mirz-2}
\begin{align}
A^{0\,JT}_4(\kappa_0 \dots \kappa_3) &=   \sum \limits_{i=1}^3 \sum \limits_{\underset{ m,n \in \mathbb{Z}}{m \neq n;}} \frac{(-1)^{m+n}}{4\pi^2 z_n z_m} \frac{ \prod \limits_{j = 0,i}\cos (2\pi \kappa_j z_m) \prod \limits_{j \neq 0,i}\cos (2\pi \kappa_j z_n) }{(z_m - z_n)^2} - \nonumber \\
& -\sum \limits_{m \in \mathbb{Z}}  \frac{\left(\prod \limits_{i=0}^3 \cos (2 \pi \kappa_i z_m) \right)}{2z_m^2} \left(-1 + \sum \limits_{i=0}^3 \kappa_i^2\right)
\end{align}
One can recognize the usual WP volume $\frac{V_{0,4}(\vec{l} = 2\pi i \vec{\kappa})}{2\pi^2}$ as a factor in the last term.
 
Let us comment on how one can motivate the formula \eqref{int-theor-conj} by explicit calculation in this limit. The general formula in \cite{Dunin-Barkowski:2012kbi}, representing $\omega_{g,n}$ as sums over stable graphs, says that the factors associated to the edges of the graph are of the form $\sum \limits_{r,s=0}^\infty B_{m_1, 2r, m_2,2s} \psi_\bullet^r \psi_\circ^s $,
where coefficients $B_{m,r,l,s}$ are determined by expanding the bidifferential $\omega_{0,2}$ in a suitable local coordinate $\zeta$ (it should satisfy $\br{\zeta}^{(m)}(\zeta^{(m)}) = - \zeta^{(m)}$; in our case we can take $\zeta^{(m)} = z-z_m$) when $z_1$ and $z_2$ are in the vicinity of branching points $z_{m_1}$ and $z_{m_2}$:
\begin{equation}
\omega_{0,2} = \left(\frac{\delta_{m_1, m_2}}{(z_1 -z_2)^2} + 2\pi \sum \limits_{r,s=0}^\infty \frac{B_{m_1, r, m_2,s} (z_1-z_{m_1})^r (z_2-z_{m_2})^s}{\Gamma (\frac{r+1}{2})\Gamma (\frac{s+1}{2})} \right) dz_1dz_2
\end{equation}
In our case of standard bidifferential and $z_m = \frac{1}{4}(2m+1)$ the coefficients are computed to be
\begin{equation}
\sum \limits_{r,s=0}^\infty B_{m_1, 2r, m_2,2s} \psi_\bullet^r \psi_\circ^s = \frac{1}{\Psi} \sum \limits_{n=1}^\infty \frac{(2n)!}{n!} \frac{\Psi^n}{(m_1 - m_2)^{2n}} \delta_{m_1 \neq m_2},\, \Psi =\psi_\bullet + \psi_\circ
\end{equation}
To get a similar representation to finite $p$ case, with <<fusion coefficients>> entering the vertices, this series should be represented as an integral of the form $\int \limits_{0}^1 d\kappa \cos (2\pi \kappa z_{m_1}) \cos (2\pi \kappa z_{m_2}) \dots$ (this is the ``continuous'' analogue of $\sum \limits_{k=1}^p \sin \pi k m_1 b^2 \sin \pi k m_2 b^2 \dots$).  This amounts to computing inverse Fourier transform of this series with respect to $(m_1 - m_2)$, for which we use
\begin{equation}
\sum \limits_{m \in \mathbb{Z}, m \neq 0} \frac{\cos (\pi \kappa m)}{m^{2n}} = \frac{(2\pi i)^{2n}}{(2n)!} B_{2n} \left(\frac{\kappa}{2}\right)
\end{equation}
Then, the same series as in \eqref{int-theor-conj} appears, with suitable scaled parameter
\begin{equation}
\sum \limits_{m \in \mathbb{Z}}\frac{\cos (\pi \kappa m)}{\Psi} \sum \limits_{n=1}^\infty \frac{(2n)!}{n!} \frac{\Psi^n}{m^{2n}} \delta_{m \neq 0} =  \sum \limits_{n=1}^\infty \frac{(-4\pi^2 \Psi)^{n-1}}{n!} B_{2n}\left(\frac{\kappa}{2}\right)
\end{equation}

\section{On ground ring operator insertions in TR formulation} \label{sec-gr-corr}
\subsection{Conjecture}
Finally, we describe one more reason one might be interested in a reformulation of MS matrix dual in terms of spectral curve \eqref{sc2}. While the supposed analogues of correlators with tachyons are more or less understood, it was not clear in other approaches whether the amplitudes including operators other than tachyons are calculable in the dual theory. We believe that, for example, for the amplitudes involving ground ring the answer might exist in this approach. 

Let us first motivate our conjecture. Consider the $n$-point amplitude with the insertion of the simplest ground ring field $O_{1,1}$ --- the unit operator (for example, let all other operators be tachyons). Obviously it should be the same as the $(n-1)$-point amplitude, that is calculated by applying \eqref{w-check-to-A} to $\check{\omega}_{g,n-1}$. Can we find a linear transformation of the $n$-differential $\check{\omega}_{g,n}$ with respect to $n$-th argument that would produce this result? 

There is a general known property of TR that hints at the answer --- the so-called ``dilaton equation''. In our case it reads
\begin{equation}
\left(\sum \limits_m \underset{z=\zeta_m}{\text{Res}} = - \underset{z=\infty}{\text{Res}} \right)\left(\Phi_1(z) \check{\omega}_{g,n}(z,z_1 \dots z_{n-1}) \right) = (2g-2+n-1) \check{\omega}_{g,n-1}(z_1 \dots z_{n-1})
\end{equation}
where $\Phi = \int \check{y}\,d\check{x}$. Using trigonometric form of Chebyshev polynomials
\begin{equation}
\Phi_1(z=\cos \theta) = -(2p+1) \int \cos (2\theta) \sin ((2p+1)\theta) d\theta = \frac{2p+1}{2}\left(\frac{T_{2p+3}(z)}{2p+3} + \frac{T_{2p-1}(z)}{2p-1} \right)
\end{equation}
Then, up to a factor that depends on $g$ and $n$ (but does not depend on parameters of other insertions) transformation $\sum \limits_m \underset{z=\zeta_m}{\text{Res}}\,\Phi_1(z)\cdot ... $ acts as an insertion of unit operator for any correlator. Note that $\Phi_1$ is a polynomial of degree $2p+3$; if we extrapolate the correspondence between the degree of the polynomial and the Liouville momentum noted in the footnote after \eqref{w-check-to-A}, this degree corresponds to unit operator $V_{1,1}$ in the Liouville sector. This is also reassuring.

On the example of 4-point amplitude with 3 tachyon insertions, we can see explicitly how this transformation reduces it to three-point correlator. Taking residue with $\Phi_1$ is the same as taking the sum $\check{A}^0_4(1, \dots) + \check{A}^0_4(-1, \dots)$, with $\check{A}^0_4$ defined by \eqref{04-answer}. When we change $k_1 \to -k_1$, Verlinde fusion numbers defined in \eqref{verl} simply change sign. This means that in this sum, terms corresponding to the trivalent stable graph with two vertices disappear. The other terms almost cancel each other, because
\begin{equation}
    \frac{V^{p}_{0,4}(4\pi b P_{1,-1},\dots) - V^{p}_{0,4}(4\pi b P_{1,1},\dots) }{2\pi^2\cdot 2} = 2b^2
\end{equation}
with the dependence on $k_{2,3,4}$ remaining only in the fusion factor, that in this case reduces to the three-point one.

In fact, since there is a more general formula
\begin{equation}
    \frac{V^{p}_{0,4}(4\pi b P_{1,-k},\dots) - V^{p}_{0,4}(4\pi b P_{1,k},\dots)}{2\pi^2\cdot 2k} = 2b^2
\end{equation}
similar thing happens for four-point correlator if we consider a transformation
\begin{equation} \label{w-check-to-gr}
\underset{z=\infty}{\text{Res}}\,\left(\Phi_k(z)\cdot ...\right) ,\,\Phi_k(z) = \frac{1}{2 k} \left(\frac{T_{2(p-k)+1}}{2p-2k+1} + \frac{T_{2(p+k)+1}}{2p+2k+1} \right)(z)
\end{equation}
If we take the residue with respect to other 3 variables, multiplied by Chebyshev polynomials according to \eqref{w-check-to-A},  we are left with is just the four-point fusion constant $\mathcal{N}_{k k_2k_3k_4}^{(0)}$ up to some constant. This is precisely the answer we expect for a four-point correlator of $O_{1,k}$ and three tachyons. 

For this more general transformation, using the explicit expressions calculated in the present paper, one can also check the following:
\begin{enumerate}
    \item 
    \begin{equation}
    \underset{z_1 = \infty}{\text{Res}} \dots\underset{z_5 = \infty}{\text{Res}} \left(\Phi_{k_1}(z_1) \Phi_{k_2}(z_2) \left(\prod \limits_{i=3}^5 \frac{T_{2(p-k_i)+1}(z_i)}{2(p-k_i)+1}\right) \check{\omega}_{0,5}(z_1 \dots z_5) \right) = \# \mathcal{N}^{(0)}_{k_1 \dots k_5}
    \end{equation}
    which is what is expected from a five-point correlator with two ground ring insertions. To check this explicitly, one can use the relation
    \begin{equation} \label{v05-v04-con}
\frac{V^{p}_{0,5}(4\pi b P_{1,-k},\dots) - V^{p}_{0,5}(4\pi b P_{1,k},\dots)}{4\pi^4\cdot 2k} = \frac{4b^2 V^{p}_{0,4}(\dots) }{2\pi^2} - 2b^6 (k^2-1)
\end{equation}
    \item 
    \begin{equation}
    \text{Res} \left(\Phi_{k_1}(z_1) \check{\omega}_{1,1}(z_1) \right) = \# \mathcal{N}^{(1)}_{k_1}
    \end{equation}
    which is also the correct answer for ground ring average on the torus (see \cite{Artemev:2022sfi}).
\end{enumerate}
All of this brings us to the conjecture: the transformation \eqref{w-check-to-gr} of topological recursion data has something to do with computing the correlators with ground ring insertions.

This conjecture still might need some refining. First of all, after this transformation a factor dependent on $g,n$ analogous to the one in dilaton equation appears, that needs to be understood.  Second, for a non-trivial check of this proposal it would be nice to match some examples of the correlators involving integrated tachyons, e.g. the amplitude with 1 ground ring operator and 4 tachyons. This will be discussed in what follows.

\subsection{Check against HEM}
The conjectured formula for correlator of $4$ tachyons and ground ring operator that follows from \eqref{w-check-to-gr} can be derived using \eqref{v05-v04-con}. It reads
\begin{align} \label{gr-5pf-tr}
\frac{1}{4b^4}\text{Res} \left(\Phi_{k_1}(z_1) \left(\prod \limits_{i=2}^5 \frac{T_{2(p-k_i)+1}(z_i)}{2(p-k_i)+1}\right) \check{\omega}_{0,5}(z_1 \dots z_5) \right) = \nonumber \\
=\mathcal{N}_{k_1 \dots k_5}^{(0)}\left( \frac{4b^2 V^{p}_{0,4}(4\pi b P_{1,-k_2},\dots 4\pi b P_{1,-k_5}) }{2\pi^2} - 2b^6 (k_1^2-1) \right)  +\frac{2b^2}{2}  \sum \limits_{i<j,2}^5 \sum \limits_{n=1}^{2p} G(n) \mathcal{N}^{(0)}_{k_i k_j n}  \mathcal{N}^{(0)}_{n \cdot \cdot \cdot}
\end{align} 

In the worldsheet approach, the correlator is built as follows. Ground ring operator can be inserted in an arbitrary point on the surface --- the result should be 
independent of this point because of \eqref{posind}. We are left with one integration over the surface: one of the tachyons has to be in ``integrated'' form to satisfy the ghost number constraint. We are left with the following integral $\left\langle \mathcal{O}_{1,k_1}(x) \int d^2z\,\mathcal{U}_{1,k_2}(z)\,  \mathcal{W}_{a_1}(x_1) \mathcal{W}_{a_2}(x_2) \mathcal{W}_{a_3}(x_3)\right\rangle $.

Applying HEM to the integrated tachyon then leads to the following expression (for a more thorough explanation see \cite{Artemev:20225pf}, where this correlator appeared as one of the terms in the expression for 5-point tachyon amplitude)
\begin{align}
    &\left\langle \mathcal{O}_{1,k_1}(x) \int d^2z\,\mathcal{U}_{1,k_2}(z)\,  \mathcal{W}_{a_1}(x_1) \mathcal{W}_{a_2}(x_2) \mathcal{W}_{x_3}(\infty)\right\rangle_{\text{HEM}} \sim \nonumber \\
    &\sim \sum \limits_{l  = - (k_1-1):2}^{k_1 - 1} \sum \limits_{m=-(k_2-1):2}^{k_2-1}  \left(q_{0,-m}^{1,k_2} (a_1) + q_{0,-m}^{1,k_2}(a_2) + q_{0,-m}^{1,k_2}(a_3) \right)  - \nonumber \\
    &- 2i P_{1,k_2} k_1 k_2 +\sum \limits_{s = |k_2 - k_1|+1:2}^{k_2 + k_1 - 1} q_{0,s-k_1}^{(1,k_2)}(a_{1,k_1}) \cdot s \label{gr-ring-corr-hem}
\end{align}
The last term is the boundary contribution in the vicinity of $z=x$ that would follow from \eqref{gr-gr-ope}. 

Let us compare the HEM answer versus the one that follows from the ``conjecture'' \eqref{gr-5pf-tr}. 
Consider ``generic'' values of three tachyon parameters $k_3, k_4, k_5$ (the ones for which the HEM calculation should apply); ``genericness'' implies the following for the fusion numbers
\begin{equation}
\mathcal{N}^{(0)}_{k_1 k_2 a_3 \dots a_5} = k_1 k_2;\,\mathcal{N}^{(0)}_{k_i a_j a_l a} = k_i;\,\mathcal{N}^{(0)}_{k \,a \,a+lb/2} =
\begin{cases}
  1,\, l =-k+1:2:k-1 \\
  0,\,\text{else}
\end{cases}
\end{equation}
Abusing notation, we write $a$ instead of $k$ in the argument of $\mathcal{N}$ for ``generic'' parameters. The HEM answer for these values, in proper normalization, can be written as
\begin{align}
&\left\langle \mathcal{O}_{1,k_1}(x) \int d^2z\,\mathcal{U}_{1,k_2}(z)\,  \mathcal{W}_{a_1}(x_1) \mathcal{W}_{a_2}(x_2) \mathcal{W}_{x_3}(\infty)\right\rangle_{\text{HEM}} = \nonumber \\
& =4b^2\left(k_1 \alpha_{0,4}(k_2|a_3,a_4,a_5) -
\underbrace{4 b\sum \limits_{s = |k_2 - k_1|+1:2}^{k_2 + k_1 - 1} q_{0,s-k_1}^{(1,k_2)}(a_{1,k_1}) \cdot s}_{(1)} \right)
\end{align}
where
\begin{equation}
   \alpha_{0,4}(k_2|a_3,a_4,a_5) = \frac{V^p_{0,4}(4\pi b P_{1,-k_2},\dots)}{2\pi^2} \mathcal{N}^{(0)}_{k_2 \dots k_5} + \sum \limits_{i=3,4,5} \sum \limits_{l=-k_2+1:2}^{k_2-1}  G(k_i+l)
\end{equation}
\eqref{gr-5pf-tr}, on the other hand, reads
\begin{align}
&\frac{1}{4b^4}\text{Res} \left(\Phi_{k_1}(z_1) \left(\prod \limits_{i=2}^5 \frac{T_{2(p-k_i)+1}(z_i)}{2(p-k_i)+1}\right) \check{\omega}_{0,5}(z_1 \dots z_5) \right)  = \nonumber \\
&=4b^2 \left(k_1 \frac{V^{p}_{0,4}(4\pi b P_{1,-k_2},\dots) }{2\pi^2} + \sum \limits_{i=3,4,5} \sum \limits_{l=-k_2+1:2}^{k_2-1}  \underbrace{k_1}_{\mathcal{N}_{k_1m \bullet \circ}} G(k_i+l) - b^4\frac{k_1 k_2 (k_1^2-1)}{2}  +\right. \label{diff0} \\
&\left. +\frac{1}{2} \sum \limits_{i=3,4,5} \sum \limits_{l=-k_2+1:2}^{k_2-1} \sum \limits_{m=-k_1+1:2}^{k_1-1} \left(G(k_i+m+l)-G(k_i+l) \right)  \right) \label{diff}
\end{align}
The first two terms coincide with $k_1 \alpha_{0,4}(k_2|a_3,a_4,a_5) $. Recall that  $$ G(m) \equiv   \left(\frac{b^4}{12} + \frac{2}{3} + b^4 (p+\frac{1}{2}-m)^2 - 2b^2 |p+\frac{1}{2}-m|. \right)$$ In the sum \eqref{diff}, quadratic terms can be simplified to
\begin{equation}
\frac{b^4}{2} \sum \limits_{i=3,4,5} \sum \limits_{l=-k_2+1:2}^{k_2-1} \sum \limits_{m=-k_1+1:2}^{k_1-1} \left((p+\frac{1}{2}-k_i-m-l)^2-(p+\frac{1}{2}-k_i-l)^2 \right) = \frac{b^4}{2} \frac{3}{3} k_1 k_2(k_1^2-1)  
\end{equation}
which cancels the last term in \eqref{diff0}. The linear terms can be represented as
\begin{align}
&-b^2 \sum \limits_{i=3,4,5} \sum \limits_{l=-k_2+1:2}^{k_2-1} \sum \limits_{m=-k_1+1:2}^{k_1-1} \left(|p+\frac{1}{2}-k_i-m-l|-|p+\frac{1}{2}-k_i-l| \right) 
 = \nonumber \\
 & =\underbrace{2b \sum \limits_{i=3,4,5}    \sum \limits_{l  = - (k_1-1):2}^{k_1 - 1} \sum \limits_{m=-(k_2-1):2}^{k_2-1}\left( q_{0,-m}^{(1,k_2)} (a_i)  -q_{0,-m}^{(1,k_2)}(a_i+l b/2)  \right)}_{(2)}
\end{align}
So, the difference of the two answers is (1) minus (2).

\subsection{Discussion}
 The two answers do not match, although the difference is quite simple. Let us discuss what this might imply if we take the conjecture \eqref{gr-5pf-tr} seriously.
 
There are some reasons to doubt whether the HEM expression actually computes the minimal string amplitude. First, naively there seems to be an alternative way to compute it: using independence on $x$, ground ring can be moved arbitrarily close to points of insertion of other tachyons to be able to perform OPE using \eqref{gr-tach-ope} and reduce the correlator to sum of four-point amplitudes. However, the result depends on what point one moves the ground ring operator towards and does not coincide with \eqref{gr-ring-corr-hem}. E.g. if we perform OPE of $O$ and $W_{a_1}$, the answer will be
\begin{equation} \label{o-w-ope-ans}
 \sum \limits_{l  = - (k_1-1):2}^{k_1 - 1} \sum \limits_{m=-(k_2-1):2}^{k_2-1}\left(q_{0,-m}^{(1,k_2)}(a_1+l b/2)  + + q_{0,-m}^{1,k_2}(a_2) + q_{0,-m}^{1,k_2}(a_3) \right)  - 2i P_{1,k_2} k_1 k_2
\end{equation}
and the difference from the HEM answer is of the form
\begin{equation}
     \sum \limits_{s = |k_2 - k_1|+1:2}^{k_2 + k_1 - 1} q_{0,s-k_1}^{(1,k_2)}(a_{1,k_1}) \cdot s + \sum \limits_{l  = - (k_1-1):2}^{k_1 - 1} \sum \limits_{m=-(k_2-1):2}^{k_2-1}\left( q_{0,-m}^{(1,k_2)} (a_1)  -q_{0,-m}^{(1,k_2)}(a_1+l b/2)  \right) \label{gr-hem-ope-diff}
\end{equation}
Asymmetry with respect to permutation of tachyon parameters (which is a property expected of string amplitudes) is concerning. In \eqref{gr-ring-corr-hem} this symmetry is also not evident (although the same can be said about \eqref{4-pf-HEM}). 

Another concern is related to what was discussed in section \ref{disc-analysis} (discontinuity analysis).
If the property stated in the end of section \ref{disc-analysis} is valid in this generalized sense, for a five-point correlator with ground ring and tachyons there should be additional non-analytic terms other than the ones present in \eqref{gr-ring-corr-hem}. Namely, \eqref{gr-ring-corr-hem} has the terms with discontinuities associated to a degeneration limit when ground ring and integrated tachyon are on the different smooth components of a degenerate curve (these are of the form $|a_i - \frac{Q}{2} + \frac{m b}{2}|,\, m \in -(k_2-1):2:k_2-1$), but does not account for the case when they are on the same component (these should be of the form $|a_i - \frac{Q}{2} + \frac{m b}{2} + \frac{l b}{2}|,\,l \in -(k_1-1):2:(k_1-1),\, m \in -(k_2-1):2:k_2-1$). Such terms, on the other hand, are present in \eqref{o-w-ope-ans}. Both of these are necessary if we want the correlator to be symmetric under permutation of 4 tachyon operators.

 Unlike the HEM expression, the answer \eqref{gr-5pf-tr} is manifestly symmetric under permutations of $k_2 \dots k_5$ and predicts expected discontinuities.  An important question then is to resolve the conflict between different answers and compare \eqref{gr-5pf-tr} (as well as HEM prediction) with the integral $\left\langle \mathcal{O}_{1,k_1}(x) \int d^2z\,\mathcal{U}_{1,k_2}(z)\,  \mathcal{W}_{a_1}(x_1) \mathcal{W}_{a_2}(x_2) \mathcal{W}_{a_3}(x_3)\right\rangle $ we study. 
 
 One more speculation on the relation between different answers is as follows. As mentioned before, apparent possibility to apply $O W$ OPE is in conflict with $x$-independence of the correlator. One option is that the $x$-independence is valid on the open part of moduli space, but breaks down on the boundary, on which the correlator takes a different value. Perhaps the symmetric tachyon amplitude \eqref{gr-5pf-tr}  can be obtained by adding to the integral over open part of moduli space, given by expression akin to \eqref{gr-ring-corr-hem}, some contribution from boundary strata of $\mathcal{\br{M}}_{0,5}$ which would look similar to \eqref{o-w-ope-ans}.

\section{Conclusions}
We finish with listing some directions for further studies.
\begin{itemize}
    \item It is of interest to prove  the diagrammatic expansion that computes $\check{A}^g_n$ that is hinted at by obtained expressions, or to find a different way to streamline the calculation of amplitudes in this approach. Perhaps one can adapt the universal $x-y$ swap formula \cite{Alexandrov:2022ydc} for this purpose.
    \item We plan to study further the 5-point amplitude in genus zero with ground ring and 4 tachyon insertions and address the tension between different analytic approaches from ``first principles'' --- by  calculating it numerically. We need to perform the same number of integrations as for well-studied example of $4$-point tachyon amplitude (see e.g. \cite{alesh2016}), but the problem incorporates 5-point conformal blocks and correlator of descendants, which adds some difficulty. Another interesting example that should have less subtleties for numerical calculations is the torus two-point amplitude of ground ring + tachyon (a similar calculation was done for $c=1$ string in \cite{Balthazar:2017mxh}). We plan to perform these calculations in the near future \cite{inprep}.
    \item The main observation of this paper raises a question of whether there is an analogous simple way to match correlators in two approaches for the general $(p,q)$ minimal string case. This is a much more subtle issue than for Lee-Yang series: even for three-point amplitudes on the sphere the answer (fusion coefficient) can not be simply reproduced by either the usual or $x-y$ dual spectral curve; instead it is given by the product of the two answers \cite{Marshakov:2024lvk}. On the other hand, it is very easy to imagine a natural generalization of \eqref{int-theor-conj-final} to this case, by replacing the fusion factors with the appropriate ones for $(p,q)$ model and extending the sum over $k_e$ to the lattice of all Liouville momenta of the form $P_{m,n}$. Results of the current paper suggest that ramification points are naturally associated with primary fields in the matter sector; perhaps there is a more convenient TR formulation with $(p-1)(q-1)$ ramification points instead of the standard one coming from the two-matrix model.
\end{itemize}

\section*{Acknowledgements}
The author is grateful to Maxim Kazarian, Igor Chaban, Alexey Litvinov, Vladimir Belavin, Dmitry Khromov, Andrei Marshakov, Lorenz Eberhardt and Wayne Weng for stimulating discussions and correspondence. He is also thankful to Adam Levine and Sergey Shadrin for discussions and pointing out some typos in the previous version. He also would like to thank the organizers of 2024 Les Houches summer school ``Quantum Geometry'', where this work was initiated, for the productive environment and the organizers of Kyushu IAS-iTHEMS workshop ``Non-perturbative methods in QFT'' for the opportunity to present preliminary results of this work at the poster session. This work has been supported by the Russian Science Foundation under the grant 23-12-00333.
\appendix

\section{Cancellation of ``on-shell poles'' from descendants} \label{app-on-shell}
The statement we prove here is as follows: for the product of elliptic four-point blocks of the form appearing in the integrand for four-point function
\begin{equation}
\mathcal{H}^{L,b}(P,p_i\mid q) \mathcal{H}^{M,i b}(P_M,ip_i\mid q) \label{prod}
\end{equation}
(the relation between $\mathcal{F}$ and $\mathcal{H}$, as well as between $q$ and $z$ can be found in \cite{Zamolodchikov:1987rec2})  $k$-th order coefficient in $q$ ($k>0$) is proportional to $P^2 + P_M^2 + k$. This means that the potential ``on-shell pole'' coming from $\int d^2q\,|q|^{2(P^2 + P_M^2 + k -1)} \sim \frac{1}{P^2 + P_M^2 + k}$ is cancelled. The same apparently is valid for $z$-blocks $\mathcal{F}$.

Expansion coefficients in $q$ ($\mathcal{H}^{L/M}(P,p_i\mid q) = \sum \limits_{k=0}^\infty q^k C_{L/M}(k,P^2)$) for these blocks obey Zamolodchikov recursion \cite{Zamolodchikov:1984rec2, Zamolodchikov:1987rec2}
\begin{equation}
C_L(k,P^2) = \delta_{k,0} + \sum \limits_{rs \leq k} C_L(k-rs, p_{r,s}^2 + r s) \frac{\alpha_{r,s}}{P^2 - P_{r,s}^2} 
\end{equation}
\begin{equation}
C_M(k,P_M^2) = \delta_{k,0} + \sum \limits_{rs \leq k} C_M(k-rs, -P_{r,s}^2) \frac{\alpha_{r,s}}{P_M^2 + P_{r,s}^2 + rs} 
\end{equation}
$\alpha_{r,s}$ are factors dependent on external dimensions and central charge; it turns out that they coincide for $\mathcal{H}^L$ and $\mathcal{H}^M$ when the block parameters are related as in (\ref{prod}). 
In both formulas by $P_{r,s}$ we mean degenerate momenta for Liouville sector. To express the recursion in this terms we use $P_{r,s}^{2}(ib) = -P_{r,-s}^2(b) = - P_{r,s}^2(b) + rs$. 

$k$-th coefficient in the product (\ref{prod}) is a polynomial in $\alpha_{r,s}$ containing such that for each monomial $\alpha_{r_1, s_1} \dots \alpha_{r_m, s_m}$ $\sum r_i s_i = k$. In this polynomial, coefficient before $\alpha_{r_1, s_1} \dots \alpha_{r_m, s_m}$ is given by a sum over permutations of the form
\begin{align}
&\sum \limits_{A \cup B = \lbrace 1 .. m \rbrace} \sum \limits_{\sigma_A, \sigma_B} \left(\frac{1}{P^2 + a_{\sigma_A(1)} } \frac{1}{b_{\sigma_A(1)} + a_{\sigma_A(2)} } \dots \frac{1}{b_{\sigma_A(|A|-1)} + a_{\sigma_A(|A|)} } \right) \times \nonumber \\
& \times\left(\frac{1}{P_M^2 + b_{\sigma_B(1)} } \frac{1}{a_{\sigma_B(1)} + b_{\sigma_B(2)} } \dots \frac{1}{a_{\sigma_B(|B|-1)} + b_{\sigma_B(|B|)} } \right)
\end{align}
where $\sigma_A, \sigma_B$ are permutations of the sets $A,B$; $a_i \equiv -P_{r_i,s_i}^2, b_i \equiv P_{r_i,s_i}^2 + r_i s_i$. Now we prove that this sum is proportional to
\begin{equation}
P^2 + P_M^2 + \sum \limits_{i=1}^m a_i + \sum \limits_{i=1}^m b_i = P^2 + P_M^2 + \sum \limits_{i=1}^m r_i s_i =P^2 + P_M^2 + k
\end{equation}
It is equivalent to show the following: renaming additionally $P^2 = b_0$, $P_M^2 = a_0$, for any $m$ sum over permutations $\sigma$ of $\lbrace 0 .. m\rbrace$
\begin{equation}
\sum \limits_{\sigma} \prod \limits_{i=0}^{m-1} \frac{1}{a_{\sigma(i)} + b_{\sigma(i+1)}} \sim \sum \limits_{i=0}^m \left(a_i + b_i\right)
\end{equation}
This sum over permutations can be rewritten as a sum over $m!$ orbits of cyclic permutation group. For each orbit, the summand is
\begin{equation}
\sum \limits_{\text{cycle}} \frac{a_{\sigma(m)} + b_{\sigma(0)}}{a_{\sigma(m)} + b_{\sigma(0)}} \prod \limits_{i=0}^{m-1} \frac{1}{a_{\sigma(i)} + b_{\sigma(i+1)}} = \frac{1}{(a_{\sigma(1)} + b_{\sigma(2)}) \dots (a_{\sigma(m)} + b_{\sigma(0)})} \sum \limits_{\text{cycle}} \left( a_{\sigma(m)} + b_{\sigma(0)} \right)
\end{equation}
Denominator is invariant under cyclic permutations, so it can be factored away; the remaining sum of $m+1$ terms is precisely  $\sum \limits_{i=0}^m \left(a_i + b_i\right)$. Thus, we proved the statement.

\section{Three-point amplitudes via resonance transformations: example $(2,11)$} \label{app:211}
Here we provide an example of how the procedure described in \ref{reson-tr} works. Third derivatives with respect to $\tau$ are computed using simple TR formula \eqref{kp-deriv} for $\frac{\pd^n F}{\pd t_{k_1} \dots \pd t_{k_n}}$, ``resonance transformations'' \eqref{taudef} and
\begin{align*}
\frac{\pd^3 \mathcal{F}}{\pd \tau_{k_1} \pd \tau_{k_2} \pd \tau_{k_3}} &= \sum \limits_{m_{1,2,3} = 1}^{p}\frac{\pd^3 \mathcal{F}}{\pd t_{m_1} \pd t_{m_2} \pd t_{m_3}} \frac{\pd t_{m_1}}{\pd \tau_{k_1}} \frac{\pd t_{m_2}}{\pd \tau_{k_2}} \frac{\pd t_{m_3}}{\pd \tau_{k_3}} + \sum \limits_{m = 1}^{p}\frac{\pd \mathcal{F}}{\pd t_{m}} \frac{\pd^3 t_{m}}{\pd \tau_{k_1} \pd \tau_{k_2} \pd \tau_{k_3}} + \\
& + \sum \limits_{m_{1,2} = 1}^{p} \frac{\pd \mathcal{F}}{\pd t_{m_1} \pd t_{m_2}}\left( \frac{\pd t_{m_1}}{\pd \tau_{k_1}} \frac{\pd^2 t_{m_2}}{\pd \tau_{k_2} \tau_{k_3}} + \frac{\pd t_{m_1}}{\pd \tau_{k_2}} \frac{\pd^2 t_{m_2}}{\pd \tau_{k_1} \tau_{k_3}} + \frac{\pd t_{m_1}}{\pd \tau_{k_3}} \frac{\pd^2 t_{m_2}}{\pd \tau_{k_2} \tau_{k_1}}  \right)
\end{align*}
We also use the following formulas for derivativeswith respect to KP times in the unstable cases \cite{Marshakov:2024lvk}
\begin{equation}
\frac{\pd \mathcal{F}}{\pd t_{k}} = \underset{z=\infty}{\text{Res}} \frac{x^{p-k+1/2}\,y\,dx}{p-k+1/2},\,\frac{\pd \mathcal{F}}{\pd t_{k_1} \pd t_{k_2}} = \underset{z_1=\infty}{\text{Res}}\, \underset{z_2=\infty}{\text{Res}} \frac{x(z_1)^{p-k_1+1/2}x(z_2)^{p-k_2+1/2}\,\omega_{0,2}(z_1,z_2)}{4(p-k_1+1/2)(p-k_2+1/2)}
\end{equation}
Then, $A^0_3(k_1, k_2, k_3)$ given by \eqref{agndef} up to overall normalization reads (we write these derivatives in matrix form, where $k_1, k_2 = 1..5$ are matrix indices and $k_3$ is the number of the matrix)
\begin{align}
\begin{pmatrix}
 u_0^7 & \frac{5 u_0^6}{8} & 0 & -\frac{u_0^4}{8} & 0 \\
 \frac{5 u_0^6}{8} & u_0^5 & \frac{3 u_0^4}{4} & 0 & -\frac{u_0^2}{2} \\
 0 & \frac{3 u_0^4}{4} & u_0^3 & \frac{u_0^2}{2} & 0 \\
 -\frac{u_0^4}{8} & 0 & \frac{u_0^2}{2} & u_0 & 1 \\
 0 & -\frac{u_0^2}{2} & 0 & 1 & \frac{1}{u_0} \\
\end{pmatrix},
\begin{pmatrix}
 \frac{5 u_0^6}{8} & u_0^5 & \frac{3 u_0^4}{4} & 0 & -\frac{u_0^2}{2} \\
 u_0^5 & \frac{11 u_0^4}{8} & u_0^3 & \frac{u_0^2}{2} & 0 \\
 \frac{3 u_0^4}{4} & u_0^3 & u_0^2 & u_0 & 1 \\
 0 & \frac{u_0^2}{2} & u_0 & 1 & \frac{1}{u_0} \\
 -\frac{u_0^2}{2} & 0 & 1 & \frac{1}{u_0} & \frac{1}{u_0^2} \\
\end{pmatrix},
\begin{pmatrix}
 0 & \frac{3 u_0^4}{4} & u_0^3 & \frac{u_0^2}{2} & 0 \\
 \frac{3 u_0^4}{4} & u_0^3 & u_0^2 & u_0 & 1 \\
 u_0^3 & u_0^2 & u_0 & 1 & \frac{1}{u_0} \\
 \frac{u_0^2}{2} & u_0 & 1 & \frac{1}{u_0} & \frac{1}{u_0^2} \\
 0 & 1 & \frac{1}{u_0} & \frac{1}{u_0^2} & \frac{1}{u_0^3} \\
\end{pmatrix},\nonumber \\
\begin{pmatrix}
 -\frac{u_0^4}{8} & 0 & \frac{u_0^2}{2} & u_0 & 1 \\
 0 & \frac{u_0^2}{2} & u_0 & 1 & \frac{1}{u_0} \\
 \frac{u_0^2}{2} & u_0 & 1 & \frac{1}{u_0} & \frac{1}{u_0^2} \\
 u_0 & 1 & \frac{1}{u_0} & \frac{1}{u_0^2} & \frac{1}{u_0^3} \\
 1 & \frac{1}{u_0} & \frac{1}{u_0^2} & \frac{1}{u_0^3} & \frac{1}{u_0^4} \\
\end{pmatrix},
\begin{pmatrix}
 0 & -\frac{u_0^2}{2} & 0 & 1 & \frac{1}{u_0} \\
 -\frac{u_0^2}{2} & 0 & 1 & \frac{1}{u_0} & \frac{1}{u_0^2} \\
 0 & 1 & \frac{1}{u_0} & \frac{1}{u_0^2} & \frac{1}{u_0^3} \\
 1 & \frac{1}{u_0} & \frac{1}{u_0^2} & \frac{1}{u_0^3} & \frac{1}{u_0^4} \\
 \frac{1}{u_0} & \frac{1}{u_0^2} & \frac{1}{u_0^3} & \frac{1}{u_0^4} & \frac{1}{u_0^5} \\
\end{pmatrix} \label{app:3rdtauder}
\end{align}
On the other hand, the fusion number matrices in this model are
\begin{equation}
\begin{pmatrix}
 1 & 0 & 0 & 0 & 0 \\
 0 & 1 & 0 & 0 & 0 \\
 0 & 0 & 1 & 0 & 0 \\
 0 & 0 & 0 & 1 & 0 \\
 0 & 0 & 0 & 0 & 1 \\
\end{pmatrix},
\begin{pmatrix}
 0 & 1 & 0 & 0 & 0 \\
 1 & 0 & 1 & 0 & 0 \\
 0 & 1 & 0 & 1 & 0 \\
 0 & 0 & 1 & 0 & 1 \\
 0 & 0 & 0 & 1 & 1 \\
\end{pmatrix},
\begin{pmatrix}
 0 & 0 & 1 & 0 & 0 \\
 0 & 1 & 0 & 1 & 0 \\
 1 & 0 & 1 & 0 & 1 \\
 0 & 1 & 0 & 1 & 1 \\
 0 & 0 & 1 & 1 & 1 \\
\end{pmatrix},
\begin{pmatrix}
 0 & 0 & 0 & 1 & 0 \\
 0 & 0 & 1 & 0 & 1 \\
 0 & 1 & 0 & 1 & 1 \\
 1 & 0 & 1 & 1 & 1 \\
 0 & 1 & 1 & 1 & 1 \\
\end{pmatrix},
\begin{pmatrix}
 0 & 0 & 0 & 0 & 1 \\
 0 & 0 & 0 & 1 & 1 \\
 0 & 0 & 1 & 1 & 1 \\
 0 & 1 & 1 & 1 & 1 \\
 1 & 1 & 1 & 1 & 1 \\
\end{pmatrix}
\end{equation}
It is easy to see that if we take the singular in $u_0^2$ part of matrices \eqref{app:3rdtauder} and then put $u_0 = 1$, we get fusion matrices. The regular terms, on the other hand, are not that simple.
\section{Expressions for $\check{A}^g_n$ for $(g,n)=(0,5)$ and $(1,2)$}\label{app:moreAgn}
\begin{figure}
    \centering
    \includegraphics[width=0.6\linewidth]{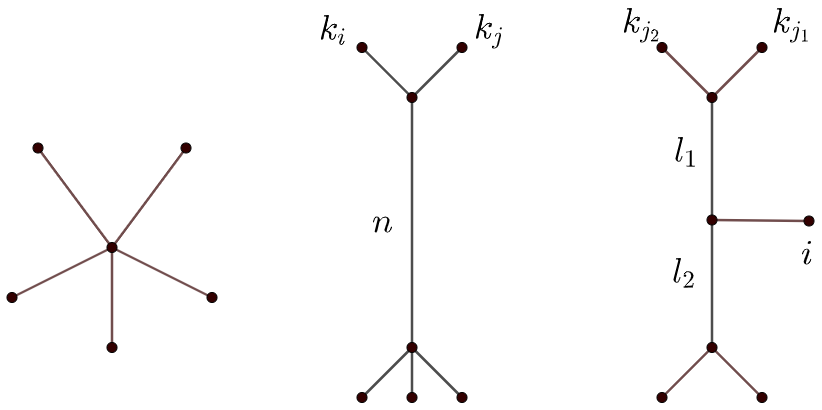}
  \caption{Stable graphs for $(g,n) = (0,5)$. All vertices have genus $g_v=0$}
    \label{fig:st-gr-05}
\end{figure}

\begin{figure}
    \centering
    \includegraphics[width=0.8\linewidth]{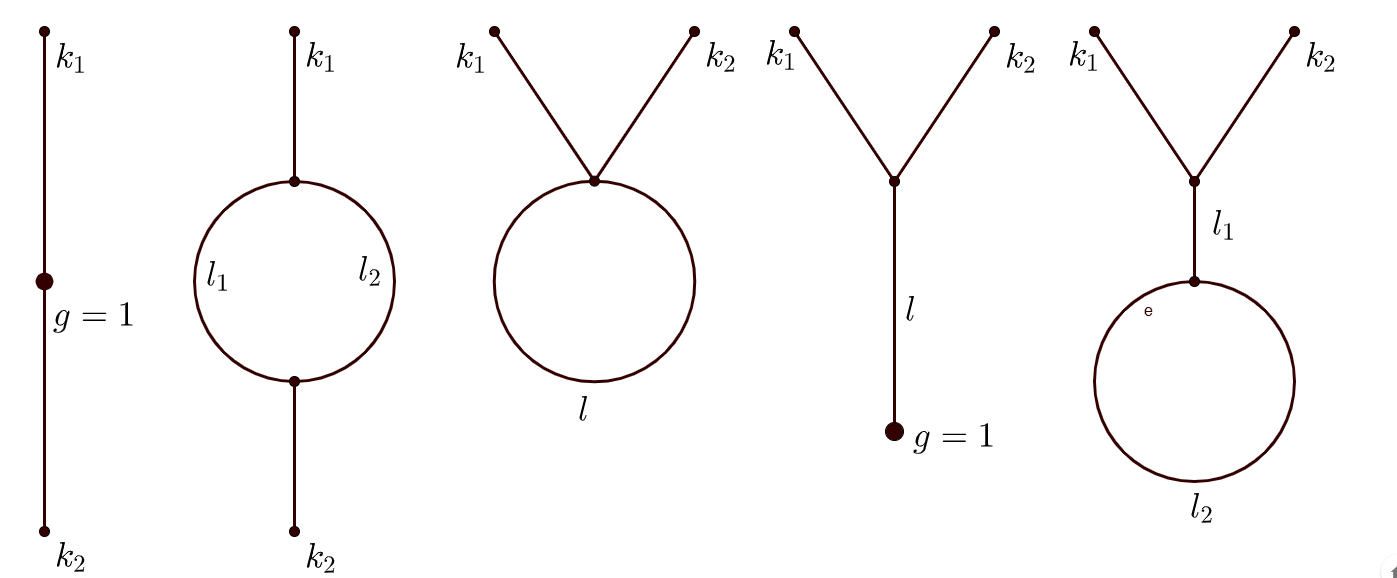}
  \caption{Stable graphs for $(g,n) = (1,2)$. We only denote the genus of vertices if it is nonzero.}
    \label{fig:st-gr-12}
\end{figure}

Calculation for $\check{A}_5^0$ gives the following answer
\begin{equation} \label{a05-tr}
4b^4\check{A}_5^0 = a_1 + a_2 + \delta a_2 + a_3 
\end{equation}
Here
\begin{align}
    a_1 = \mathcal{N}^{(0)}_{k_1 \dots k_5} \frac{V_{0,5}^b(4\pi bP_{1,-\vec{k}})}{4\pi^4}, \\
    a_2 = \frac{1}{2}  \sum \limits_{i<j,1}^5 \sum \limits_{n=1}^{2p} \mathbf{G}(n) \mathcal{N}^{(0)}_{k_i k_j n}  \mathcal{N}^{(0)}_{n \cdot \cdot \cdot} \frac{V^b_{0,4}(4\pi b P_{1,-n}, \cdot, \cdot, \cdot)}{2\pi^2},  \\
    a_3 =  \frac{1}{8} \sum \limits_{i=1}^5 \sum \limits_{j_1 < j_2; j_{1,2} \neq i} \sum \limits_{l_1,l_2}^{2p} \mathbf{G}(l_1) \mathbf{G}(l_2) \mathcal{N}^{(0)}_{k_i l_1 l_2}  \mathcal{N}^{(0)}_{k_{j_1} k_{j_2} l_1}  \mathcal{N}^{(0)}_{l_2 \cdot \cdot}
\end{align}
are contributions that can be written down using the diagrammatic technique described in section \ref{sec: feynman}; figure \ref{fig:st-gr-05} illustrates the corresponding stable graphs. However, this is not the full story --- there is an additional contribution associated with stable graphs of type $2$ of the form
\begin{align}
\delta a_2 &= \frac{b^4}{2}  \sum \limits_{i<j,1}^5 \sum \limits_{n=1}^{2p} \mathcal{N}^{(0)}_{k_i k_j n}  \mathcal{N}^{(0)}_{ n \cdot \cdot \cdot} \left(\frac{4}{15b^4} - \frac{16}{3b^2} |P_{1,-n}|^2 + 8 |P_{1,-n}|^4\right) = \nonumber \\
&= -4  \sum \limits_{i<j,1}^5 \sum \limits_{n=1}^{2p} \mathcal{N}^{(0)}_{k_i k_j n}  \mathcal{N}^{(0)}_{ n \cdot \cdot \cdot} \left(B_4(|x_n|) -2 |x_n|^2 B_2(|x_n|) \right) 
\end{align}
with
\begin{equation}
|x_n| := \frac{b^2}{2}|p+1/2-n| = b |P_{1,-n}|
\end{equation}
For torus 2-point amplitude there are 5 stable graphs, illustrated in figure \ref{fig:st-gr-12}. The corresponding decomposition of the amplitude is
\begin{equation}
8\check{A}^1_2 = b_1 + b_2 + b_3 + b_4 + b_5,
\end{equation}
where
\begin{align}
b_1 = \mathcal{N}^{(1)}_{k_1 k_2} \frac{V_{1,2}^b(P_{1,-\vec{k}})}{4\pi^4} \\
b_2 = \frac{1}{8} \sum \limits_{l_1,l_2=1}^{2p} \mathbf{G}(l_1) \mathbf{G}(l_2) \mathcal{N}^{(0)}_{k_1 l_1 l_2} \mathcal{N}^{(0)}_{k_2 l_1 l_2} \\
b_3 = \frac{1}{4} \sum \limits_{l=1}^{2p} \mathcal{N}^{(0)}_{k_1 k_2 ll}\left(\frac{V_{0,4}^b(4\pi b P_{1,-\vec{k}}, 4\pi b P_{1,-l}, 4\pi b P_{1,-l})}{2\pi^2} \mathbf{G}(l) - 16( B_4(|x_l|)-2|x_l|^2 B_2(|x_l)) \right) \\
b_4 = \frac{1}{2} \sum \limits_{l=1}^{2p} \mathcal{N}^{(1)}_l \mathcal{N}^{(0)}_{l k_1 k_2} \left(\frac{V_{1,1}^b(4\pi b P_{1,-l})}{2\pi^2}\mathbf{G}(l) - \frac{1}{3} ( B_4(|x_l|)-2|x_l|^2 B_2(|x_l))\right) \\
b_5 = \frac{1}{8} \sum \limits_{l_1,l_2 = 1}^{2p} \mathcal{N}^{(0)}_{k_1 k_2 l_1} \mathcal{N}^{(0)}_{l_1 l_2 l_2} \mathbf{G}(l_1) \mathbf{G}(l_2) 
\end{align}
We checked these results against the ones obtained in \cite{tarn2011,beltar2010} using ``resonance transformations'' and found agreement. Note that there is an error in the end of \cite{beltar2010}, where during the calculation of torus two-point amplitude authors did not account for Heaviside theta-functions in the final answer. We provide the correct formula here for reference:
\begin{align*}
&6(2p+1)^2A^1_2(k_1,k_2) = \left(-k_1 (2 p+1)+\left(p-k_2\right) \left(p-k_2+1\right)+k_1^2\right) \times \\
& \times \left(k_1 (2 p+1)+k_2 \left(2p-k_2+1\right)-k_1^2+p(p+1)+2\right)+ \\
&+\left(k_1+k_2-p-1\right) \left(k_1+k_2-p\right) \left(\left(k_1+k_2-p-2\right) \left(k_1+k_2-p+1\right)-2 p (p+1)\right) \Theta \left(p+1-k_1-k_2\right)
\end{align*}
\bibliographystyle{JHEP}
\newpage
\bibliography{mlg2}
\end{document}